\documentclass[12pt]{article}
\usepackage{amsfonts}
\usepackage{latexsym}
\usepackage{amsmath}
\usepackage{amssymb}
\usepackage{amssymb}

\usepackage[utf8x]{inputenc}
\usepackage{amsmath}
\usepackage{amssymb}
\usepackage{amsthm}
\usepackage{slashed}
\usepackage{upgreek}
\usepackage{mathrsfs}
\usepackage{bbm}
\usepackage{tikz}
\numberwithin{equation}{section}

\let\originalleft\left
\let\originalright\right
\renewcommand{\left}{\mathopen{}\mathclose\bgroup\originalleft}
\renewcommand{\right}{\aftergroup\egroup\originalright}

\newcommand\numberthis{\addtocounter{equation}{1}\tag{\theequation}}

\def\bbe{{\bf{e}}}
\font\mybb=msbm10 at 11pt

\def\bb#1{\hbox{\mybb#1}}

\def\bZ {\bb{Z}}
\def\bR {\bb{R}}

\def\bC {\bb{C}}

\renewcommand{\theequation}{\thesection.\arabic{equation}}

\def\appendix#1{\addtocounter{section}{1}\setcounter{equation}{0}
\renewcommand{\thesection}{\Alph{section}}
\section*{Appendix \thesection\protect\indent \parbox[t]{11.15cm}{#1}}
\addcontentsline{toc}{section}{Appendix \thesection\ \ \ #1}}

\newcommand{\be}{\begin{eqnarray}}
\newcommand{\ee}{\end{eqnarray}}
\newcommand{\bea}{\begin{eqnarray}}
\newcommand{\eea}{\end{eqnarray}}
\newcommand{\ba}{\begin{array}}
\newcommand{\ea}{\end{array}}

\newenvironment{frcseries}{\fontfamily{frc}\selectfont}{}

\def\ep{\epsilon}

\def\bbe{{\bf{e}}}
\font\mybb=msbm10 at 11pt

\def\bb#1{\hbox{\mybb#1}}

\def\bZ {\bb{Z}}
\def\bR {\bb{R}}

\def\bC {\bb{C}}

\begin{document}
\begin{titlepage}
\begin{center}
\vspace*{-1.0cm}
\hfill DMUS-MP-17-09 \\

\vspace{2.0cm} {\Large \bf All  Killing Superalgebras for Warped  AdS  Backgrounds} \\[.2cm]

\vspace{1.5cm}
 {\large S. Beck$^1$, U. Gran$^2$, J. Gutowski$^3$ and  G. Papadopoulos$^1$}

\vspace{0.5cm}
${}^1$ Department of Mathematics\\
King's College London\\
Strand\\
London WC2R 2LS, UK\\

\vspace{0.5cm}

${}^2$ Department of Physics\\
Division for Theoretical Physics\\
Chalmers University of Technology\\
SE-412 96 G\"oteborg, Sweden\\

\vspace{0.5cm}
$^3$ Department of Mathematics \\
University of Surrey \\
Guildford, GU2 7XH, UK \\

\vspace{0.5cm}

\end{center}

\vskip 1.5 cm
\begin{abstract}
We present  all the symmetry superalgebras $\mathfrak{g}$  of all warped AdS$_k\times_w M^{d-k}$, $k>2$, flux backgrounds in $d=10, 11$ dimensions preserving any number of supersymmetries.
First we give the conditions for  $\mathfrak{g}$  to decompose into a  direct sum of the isometry algebra of AdS$_k$ and that of the internal space $M^{d-k}$. Assuming this decomposition, we   identify all  symmetry superalgebras of AdS$_3$ backgrounds by  showing that the isometry groups of  internal spaces act transitively on  spheres.  We  demonstrate that in type II and $d=11$ theories the  AdS$_3$ symmetry superalgebras may  not be  simple and also present  all symmetry superalgebras of heterotic AdS$_3$ backgrounds. Furthermore,  we  explicitly give  the symmetry superalgebras of  AdS$_k$, $k>3$, backgrounds  and prove that they are   all  classical.

\end{abstract}

\end{titlepage}


\section{Introduction}

One way to find the symmetry superalgebra, $\mathfrak{g}=\mathfrak {g}_0\oplus \mathfrak{ g}_1$, of a product $AdS_k\times M^{d-k}$, $k>2$, background in a supergravity theory is to assume that it
 is a classical   superalgebra\footnote{The classical superalgebras are those which are simple and where the representation of $\mathfrak {g}_0$ on $\mathfrak{ g}_1$ is completely reducible \cite{kac}.} whose  even subalgebra  decomposes as ${\mathfrak g}_0={\mathfrak so}(k-1,2)\oplus \mathfrak{ t}_0$ and the dimension of the odd subspace $\mathfrak{ g}_1$ is the number of Killing spinors $N$,  where ${\mathfrak so}(k-1,2)$ is the Lie algebra of isometries of AdS$_k$ subspace. Then   $\mathfrak{t}_0$ is identified with the Lie algebra of isometries of the internal space $M^{d-k}$.  For
 $k>3$, these data together with the classification of classical superalgebras in \cite{kac, nahm} are sufficient to find all such symmetry superalgebras.

  This method based on the splitting and  classification of classical superalgebras  may be sufficient for backgrounds of the type $AdS_k\times M^{d-k}$ but that is not the case for generic warped $AdS_k\times_w M^{d-k}$  solutions.  This is because  $AdS_k$ can be written as a warped product of $AdS_m$ for any $m<k$ and so all  $AdS_k\times_w M^{d-k}$ backgrounds can be re-interpreted as $AdS_m\times_w M^{d-m}$ backgrounds \cite{strominger, desads}.  Now if all the symmetry  supergralgebras of  AdS backgrounds could be identified as described above, it would have been possible to decompose the
even subalgebra $\mathfrak{g}_0$ of the symmetry superalgebra of $AdS_k\times_w M^{d-k}$ as $\mathfrak{g}_0={\mathfrak so}(m-1,2)\oplus {\mathfrak t}'_0$.  However in all known examples this is not the case. Therefore, there must be  some conditions on the spacetime geometry required for ${\mathfrak g}_0$ to decompose as
${\mathfrak so}(k-1,2)\oplus \mathfrak{ t}_0$.  Furthermore it is not a priori obvious why one should restrict
the symmetry superalgebras of AdS backgrounds to be  classical.

 First principle computations of $\mathfrak{g}$ have also been made in the literature for many known supersymmetric solutions, see e.g.~\cite{dauriafre} and \cite{josebill, pktsuper, maxwave}. Most of these are  based on the Killing superalgebra (KSA) approach \cite{pktsuper, josesuper}  which utilizes the geometric data of the spacetime like
 the Killing spinor 1-form bilinears and the spinorial Lie derivative to define the (anti-) commutators of  $\mathfrak{g}$; the method is reviewed in section 2.  In these computations,  the geometry of the internal
 space $M^{d-k}$ is used in an essential way to determine all (anti-) commutators. Because of this, it is not apparent how to extend to general warped $AdS_k\times_w M^{d-k}$  flux backgrounds  where the geometry of the internal space $M^{d-k}$ may not be sufficiently known to find the (anti-)commutators of $\mathfrak{g}$.

In this paper we shall apply a modification of the KSA approach to identify all the symmetry superalgebras of warped AdS  backgrounds $AdS_k\times_w M^{d-k}$ with the most general allowed fluxes in 10- and 11-dimensional
supergravity theories.
 First, we find the conditions on the geometry of $AdS_k\times_w M^{d-k}$ such that ${\mathfrak g}_0$ can admit  a decomposition\footnote{One justification
 for this decomposition is AdS/CFT.  The isometry group of the AdS and internal spaces are  identified with the conformal  and R-symmetry groups of the dual theory, respectively.
 As the conformal  and R-symmetry groups of a field theory commute, ${\mathfrak g}_0$ must be a direct sum.  This is the only assumption we make. }   as ${\mathfrak g}_0={\mathfrak so}(k-1,2)\oplus \mathfrak{t}_0$.
 These conditions are expressed as vanishing conditions for certain Killing spinor bilinears or their derivatives and are stated in eqns (\ref{billI}), (\ref{billIII}) and (\ref{billII}).  We also demonstrate that the same conditions
 can be derived if one assumes that the internal space is compact without boundary and the solutions are smooth.

 Next we find that for  AdS$_3$ backgrounds the KSA decomposes as $\mathfrak{g}=\mathfrak{g}_L\oplus \mathfrak{g}_R$, where $\mathfrak{g}_L$ is associated to the left action on AdS$_3$ and $\mathfrak{g}_R$ is
   associated with the right-action on AdS$_3$ viewed locally  as a group manifold. For $N<8$ superymmetries in either the left or the right sector, the KSAs
    can be computed   from first principles. The method we use will be explained later.
For $N\geq 8$ in either the left or the right sector, we show that the isometry algebra of the internal space acts transitively on a sphere in $\mathfrak{g}_1$  and leaves a 4-form invariant.  Moreover we show that
all the structure constants of the KSA can be determined as soon as this 4-form is specified. The classification of groups acting transitively and effectively on spheres has been solved
  some time ago in \cite{ms} and  has been used \cite{simons} in the context of the Berger classification of irreducible simply connected Riemannian manifolds. Using this, {\it all  KSAs  of   AdS$_3$
  backgrounds are found  and the results are tabulated in table 2}. The table includes three series $\mathfrak{osp}(N/2\vert 2)$,  $\mathfrak{sl}(N/2\vert 2)$
and $\mathfrak{osp}^*(N/4\vert 4)$ $(N=16, 24)$\footnote{$\mathfrak{g}^*$ denotes another real form of the real superalgebra $\mathfrak{g}$.}, as well as several exceptional cases like $\mathfrak{D}(2,1,\alpha)$ ($N=8)$,
$\mathfrak{g}(3)$ $(N=14)$ and $\mathfrak{f}(4)$ $(N=16)$, where $N=\mathrm{dim}\,\mathfrak{g}_1$ is the number of supersymmetries.
It is also shown that the KSAs of AdS$_3$ backgrounds are not necessarily simple as they can exhibit
  central terms.  Though to our knowledge there are no solutions in the literature for which a central term gives rise to an effective action on the internal space.

  We also identify the KSAs of heterotic AdS$_3$ backgrounds. {\it  The results are presented in table 3}. The KSAs of heterotic AdS$_3$ backgrounds are of classical type
  and thus they do not exhibit  central generators.

  Applying the same methods to the rest of $AdS_k\times_w M^{d-k}$, $k>3$ backgrounds, we demonstrate that the KSAs are of classical type.  There is only one exception to this which
   is the KSA of maximally supersymmetric AdS$_5$ backgrounds which allows the presence of a central term.  However, as we know that the only maximally supersymmetric
   AdS$_5$ solution is locally isometric to the  $AdS_5\times S^5$ solution in IIB, one can show that the central term does not act effectively on $S^5$ and so it can be set to zero.
  The KSAs for all $AdS_k\times_w M^{d-k}$  $k>3$  backgrounds are explicitly constructed  and are related to the classification of the classical superalgebras in \cite{kac, nahm}.
{\it The list of the KSAs of $AdS_k\times_w M^{d-k}$, $k>3$ can be found in table 4 and the associated isometry algebras of the internal spaces in table 5}.

There are  two key  developments  that have allowed us to prove these results without specifying the geometry of the internal spaces.
The first is the explicit expression of the Killing spinors of
 $AdS_k\times_w M^{d-k}$ backgrounds given in \cite{mads, iibads, iiaads} for which the dependence on the AdS coordinates is manifest\footnote{It has also removed all
 the assumptions that are usually made on the form of the Killing spinors for $AdS_k\times_w M^{d-k}$ backgrounds which have been proven to be restrictive in \cite{desads}.}.   As a result, one can
determine the dependence of all 1-form Killing spinor bilinears on the AdS coordinates and to also  compute explicitly all the spinorial Lie derivatives of the Killing spinors along the isometries of  AdS. As a consequence, one can
determine the anti-commutator $\{\mathfrak{g}_1, \mathfrak{g}_1\}$  as well as all the commutators, $[\mathfrak{so}(k-1, 2), \mathfrak{g}_1]$,   of the odd generators of the KSA with
the even generators associated with isometries of AdS$_k$, $\mathfrak{so}(k-1, 2)\subseteq \mathfrak{g}_0$.
Furthermore, it is  straightforward to find the conditions (\ref{billI}), (\ref{billIII}) and (\ref{billII}) for  the even part of the superalgebra to decompose as a direct sum of the the isometry algebra of AdS space and that of the internal
 space, $\mathfrak{g}_0=\mathfrak{so}(k-1, 2)\oplus \mathfrak{t}_0$. These conditions put several  restrictions on the geometry of $AdS_k\times_w M^{d-k}$. In particular for $k>3$, they can be used
  to find the linearly independent Killing vectors of the internal space and in this way determine the dimension of $\mathfrak{t}_0$.  For $k=3$, these are sufficient to determine the
  maximal dimension of  $\mathfrak{t}_0$.

The second ingredient in our proof  is  the closure  of  KSAs for superymmetric $d=11$ and IIB  backgrounds shown in \cite{11jose, iibjose}.  We use this to demonstrate that in all cases
 the super-Jacobi identities and the explicit dependence  of the Killing spinors on the AdS coordinates are sufficient to determine the commutator $[\mathfrak{t}_0, \mathfrak{g}_1]$ from those of $\{\mathfrak{g}_1, \mathfrak{g}_1\}$ and  $[\mathfrak{so}(k-1, 2), \mathfrak{g}_1]$.  This circumvents the need to know details of the
 geometry of the internal spaces in order to find the KSAs. The remaining commutators which are those of the isometries of the internal space  can also be found after applying the super-Jacobi identities.

This paper is organized as follows. In section two, we summarize the results of \cite{mads, iibads, iiaads} and use them to find the conditions on the
geometry of  $AdS_k\times_w M^{d-k}$  such that $\mathfrak{g}_0=\mathfrak{so}(k-1, 2)\oplus \mathfrak{t}_0$.  In section 3, we classify all KSAs for AdS$_3$ backgrounds.
In section 4, we determine all KSAs for heterotic AdS$_3$ backgrounds. In sections 5, 6, 7 and 8, we show that the KSAs of AdS$_k$, $k=4,5,6$ and $7$ are classical
and give explicitly all their (anti-)commutators, respectively. In section 9, we give our conclusions. In appendix A, we demonstrate that the 1-form Killing spinor
bilinears of massive IIA supergravity leave invariant all fields of the theory. In appendix B, we give the isometries of AdS$_k$ as well as the spinorial Lie derivatives
of spinors along the AdS Killing vectors.  In appendix C, we give all the 1-form bilinears as well as the spinorial Lie derivatives of the Killing vectors
of $AdS_k\times_w M^{d-k}$ backgrounds. In appendix D, we present the construction  of KSAs for AdS$_3$ backgrounds with a low number of supersymmetries
without the use of the results of \cite{ms}.

\section{Killing  Superalgebras}

\subsection{Definition of KSAs}

Decomposing the  KSAs of  supersymmetric backgrounds $\mathfrak{g}$ into the even $\mathfrak{g}_0$ and odd subspaces $\mathfrak{g}_1$, $\mathfrak{g}= \mathfrak{g}_0\oplus \mathfrak{g}_1$, the construction proceeds as follows \cite{pktsuper, josesuper}.  $\mathfrak{g}_1$ is spanned by  the odd generators $Q_{\epsilon_{\mathbf{m}}}$ each associated to a Killing spinor $\epsilon_{\mathbf{m}}$ of the background, where $\mathbf{m}=1, \dots N$.  $\mathfrak{g}_0$  is  spanned  by the even  generators
 $V_{K_{\mathbf{m}\mathbf{n}}}$ each associated to a  1-form bilinear $K_{\mathbf{m}\mathbf{n}}$ constructed from the Killing spinors $\epsilon_{\mathbf{m}}$ and $\epsilon_{\mathbf{n}}$ as $K_{\mathbf{m}\mathbf{n}}=(\epsilon_{\mathbf{m}}, \Gamma_M \epsilon_{\mathbf{n}}) dx^M$,
where $(,)$ is a suitable $Spin(d-1,1)$-invariant inner product such that $K_{\mathbf{m}\mathbf{n}}=K_{\mathbf{n}\mathbf{m}}$.   $K_{\mathbf{m}\mathbf{n}}$ give rise to  Killing vector fields which leave all fields invariant. The (anti)commutators of KSAs are computed geometrically.  In
particular
\bea
\{Q_{\epsilon_{\mathbf{m}}}, Q_{\epsilon_{\mathbf{n}}}\}=V_{K_{\mathbf{m}\mathbf{n}}}~,~~~~[V_{K_{\mathbf{m}\mathbf{n}}}, Q_{\epsilon_{\mathbf{p}}}]=Q_{{\cal L}_{K_{\mathbf{m}\mathbf{n}}}\epsilon_{\mathbf{p}}}~,~~~[V_{K_{\mathbf{m}\mathbf{n}}}, V_{K_{\mathbf{p}\mathbf{q}}}]=V_{[K_{\mathbf{m}\mathbf{n}}, K_{\mathbf{p}\mathbf{q}}]}~,
\label{super}
\eea
where $[K_{\mathbf{m}\mathbf{n}}, K_{\mathbf{p}\mathbf{q}}]$ is the Lie commutator of two vector fields and
\bea
{\cal L}_X \epsilon=\nabla_X \epsilon+{1\over8} dX_{MN} \Gamma^{MN} \epsilon~,
\label{spinder}
\eea
 is the spinorial Lie derivative of $\epsilon$ with respect to the vector field $X$.
 It has been shown in \cite{11jose, iibjose} that closure of the KSAs  holds for all $d=11$ and IIB supersymmetric backgrounds\footnote{Closure has not been shown
  for the KSAs of massive IIA backgrounds but it is expected to hold for those as well. In appendix A, we demonstrate that the vector Killing spinor bilinears of
  massive IIA theory leave all fields of the theory invariant.}   under these operations
as the super-Jacobi identities are satisfied.
 There are several simplifications in the construction of the KSAs for AdS backgrounds which we shall explain below.
In what follows, we shall set  for simplicity $Q_{\epsilon_{\mathbf{m}}}=Q_{\mathbf{m}}$ and $V_{K_{\mathbf{m}\mathbf{n}}}=V_{\mathbf{m}\mathbf{n}}$.

\subsection{AdS Killing spinors } \label{elop}

The KSEs of all warped backgrounds  AdS$_k \times_w M^{d-k}$, $k \geq 3$, with the most general allowed fluxes can be integrated over the AdS subspace in all 10- and 11-dimensional supergravity theories \cite{mads, iibads, iiaads}.
The expression for the fluxes  depends on the theory as well as the particular AdS$_k$ background under consideration. However, the properties that will be described
 apply to all cases.  Because of this in what follows we shall focus on the metric which is universal in all theories and for the rest one should consult the references above.
 In the coordinates that the spacetime metric can be written  as
\bea
 ds^2 = 2 du \left( dr - 2 \ell^{-1} r dz - 2 r d \ln A \right) + A^2 dz^2 + A^2 e^{2 z / \ell} \delta_{a b} dx^a dx^b + g_{IJ} dy^I dy^J ,
 \label{metr}
\eea
where $(u,r,z,x^a)$ are  coordinates of the AdS subspace, $\ell$ is the radius of AdS,  $y$ are coordinates of $M$ and $A$ is the warp factor,
the expression for  Killing spinors reads
\bea
\epsilon=\epsilon_1+\epsilon_2+\epsilon_3+\epsilon_4~,
\label{ksp1}
\eea
where
\bea
\epsilon_1&=&\sigma_+~,~~~\epsilon_2=\sigma_--\ell^{-1} e^{{z\over\ell}} x^a \Gamma_{az} \sigma_--\ell^{-1} A^{-1} u \Gamma_{+z} \sigma_-~,
\cr
\epsilon_3&=&e^{-{z\over\ell}}\tau_+-\ell^{-1} A^{-1} r e^{-{z\over\ell}} \Gamma_{-z} \tau_+-\ell^{-1} x^a \Gamma_{az} \tau_+~,~~~\epsilon_4=e^{{z\over \ell}} \tau_-~,
\label{ksp2}
\eea
where the $\sigma_\pm$ and $\tau_\pm$  spinors satisfy the lightcone projections $\Gamma_\pm\sigma_\pm=\Gamma_\pm\tau_\pm=0$, and depend only on the coordinates of $M^k$
The gamma matrices have been chosen with respect to the frame
\bea
&&\bbe^+ =du~,~~\bbe^-= dr - 2 \ell^{-1} r dz - 2 r d \ln A~,~~
\cr
&&\bbe^z=A dz~,~~\bbe^a=A e^{{z\over\ell}} dx^a~,~~~\bbe^i=e^i_I dy^I~,
\eea
and    $ds^2= 2\bbe^+ \bbe^-+ (\bbe^z)^2+ \delta_{ab} \bbe^a \bbe^b+ \delta_{ij} \bbe^i \bbe^j$.
The spinors  $\sigma_\pm$ and $\tau_\pm$ satisfy some   KSEs along the internal space M$^{d-k}$  which are
\bea
D_i^{(\pm)} \sigma_\pm=D_i^{(\pm)} \tau_\pm=0~,~~~{\cal A}^{(\pm)}\sigma_\pm={\cal A}^{(\pm)}\tau_\pm=0~,
\label{kseI}
\eea
and  can be thought as the restriction of the gravitino
and dilatino KSEs of the associated supergravity theory on $M^{d-k}$, respectively,  as well as an additional algebraic KSE
\bea
\Xi^{(\pm)}\sigma_\pm=0~,~~~\Xi^{(\pm)}\tau_\pm=\mp {1\over\ell}\tau_\pm~,
\label{kseII}
\eea
which arises from the integration of KSEs on AdS.  The explicit form of (\ref{kseI}) and (\ref{kseII})  depends on the supergravity considered as well as
the specific AdS background under investigation, as the operators $D_i^{(\pm)}$, ${\cal A}^{(\pm)}$ and $\Xi^{(\pm)}$ depend on the fields.  However in all cases they take the above general form.

The $\sigma_+$ and $\tau_+$ Killing spinors lie in a complementary subspace from that of $\sigma_-$ and $\tau_-$  as they satisfy different
lightcone projections.  Moreover, it is straightforward to deduce from the algebraic KSE (\ref{kseII}) that the $\sigma_+$ Killing spinors are linearly independent
from the $\tau_+$ Killing spinors.  From this analysis, one concludes that  Killing spinors  $\epsilon_1$, $\epsilon_2$, $\epsilon_3$ and $\epsilon_4$ are all linearly independent. Later, we shall strengthen these properties using global conditions.

For AdS$_k$ $k\geq 3$,  there are elementary Clifford algebra operations that relate the $\sigma$ and $\tau$ spinors.  In particular if  $\sigma_+$ and $\tau_+$ are  Killing spinors, then $\sigma_-=A\Gamma_{-z} \sigma_+$ and $\tau_-=A\Gamma_{-z} \tau_+$ are Killing spinors.  In addition  for $k>3$, if $\sigma_+$ is a Killing spinor, then
$ \Gamma_{za} \sigma_+$ are $\tau_+$ type Killing spinors for every $a$. Moreover for $k>4$ if $\sigma_+$ is a Killing spinor,  then  $\Gamma_{ab}\sigma_+$ are $\sigma_+$ type  Killing spinors for every $a<b$.  These relations between the Killing spinors can be used to count the Killing spinors of all AdS backgrounds  \cite{mads, iibads, iiaads}. For $k=3$, it can be arranged so that for each $\epsilon$  only the $\sigma_\pm$ or the $\tau_\pm$
 spinors are non-vanishing and  also the terms proportional to $x^a$ in the expression for $\epsilon$ do not occur as the coordinates of AdS$_3$ has  $(u,r,z)$ .

 It is clear from the above that the Killing spinors of AdS backgrounds can be described in terms of multiplets. Each multiplet is determined from the choice  of $\sigma_+$.  Then the rest of
 the components of the multiplet can be constructed from $\sigma_+$ using the elementary Clifford algebra operations described above and after selecting the linearly independent spinors that
 arise from such a procedure. AdS backgrounds that preserve a  minimal amount of supersymmetry admit one such multiplet of Killing spinors while those that preserve extended supersymmetry
 admit two or more such multiplets.  We shall investigate both the properties of the Killing spinors that lie in the same multiplet as well as those of different multiplets.
 It suffices to focus on the properties of $\sigma_+$ and $\tau_+$ Killing spinors as those of $\sigma_-$ and $\tau_-$ spinors follow
 in a straightforward manner.

\subsection{1-form bilinears and decomposition of $\mathfrak{g}_0$}
\label{consection}

To investigate the conditions on the geometry of AdS$_k$$\times_w$M$^{d-k}$ backgrounds for $\mathfrak{g}_0$ to decompose as  $\mathfrak{g}_0=\mathfrak{so}(k-1,2)\oplus \mathfrak{t}_0$, we use
the 1-form bilinears of the Killing spinors (\ref{ksp2}) presented in appendix C.  These have components along the AdS subspace and components along the internal
space M$^{d-k}$.  The components along the AdS subspace span the 1-forms associated with the isometries of the AdS subspace in (\ref{metr}).
There are two kinds of components along the internal space M$^{d-k}$, those that depend on  and those that are independent from the coordinates of AdS.
The former indicate that the backgrounds have isometries along the internal space which do not commute with the isometries of AdS.  Such behavior is expected from AdS$_k$ solutions that arise as foliations of AdS$_m$ backgrounds with $k<m$. To exclude such backgrounds, it is required that all such components should vanish.  In turn, it
 is straightforward to observe that this is satisfied provided that
\bea
\langle \sigma_+, \Gamma_i \Gamma_a \sigma'_+\rangle~=0~,~~~\langle \tau_+, \Gamma_i \Gamma_z \sigma_+\rangle~=0~,
\label{billI}
\eea
where $\sigma_+'$, $\tau_+$ and $\sigma_+$ may or may not belong to the same Killing spinor multiplet, and
we have  used the relation between $\tau_-$, $\sigma_-$ and $\tau_+$, $\sigma_+$ spinors.   The inner product $\langle \cdot, \cdot\rangle$ is the real part of the standard hermitian inner product
for which all spacelike gamma matrices are hermitian.
 For $k>3$, the above two relations are equivalent as there is a relation between the $\sigma_+$ and $\tau_+$  spinors
explained in the previous section.
The remaining  1-form bilinears along the
internal space are always proportional to
$\langle \sigma_+, \Gamma_i \Gamma_z \sigma'_+\rangle~\bbe^i$,
where again $\sigma_+'$ and $\sigma_+$ may or may not belong to the same Killing spinor multiplet. These are not expected  to vanish and give rise to isometries
of the internal space.

One consequence of (\ref{billI}) is the orthogonality of $\sigma_+$ and $\tau_+$ spinors
\bea
\langle \sigma_+, \tau_+\rangle=0~.
\label{billIII}
\eea
This follows from (\ref{kseII}) after imposing (\ref{billI}).

After imposing (\ref{billI}) and (\ref{billIII}), the Killing spinor bilinears can be written as $K=K_\mu \,\bbe^\mu+ K_i \,\bbe^i$,
where $K^{\mathrm{AdS}}=K_\mu \,\bbe^\mu$ are along the AdS directions and $\tilde K=K_i \,\bbe^i$ are along the transverse directions.
$K^{\mathrm{AdS}}$ can be written as a linear combination of forms associated with Killing vector fields of the AdS space
with components  that  may depend on the coordinates of of the internal space.  Requiring that $\mathfrak{g}_0=\mathfrak{so}(k-1,2)\oplus \mathfrak{t}_0$
which implies that independently  $K^{\mathrm{AdS}}$ and $\tilde K$ are Killing and that
\bea
[K^{\mathrm{AdS}}, \tilde K]=0~,
\eea
one finds that
\bea
\parallel\sigma_+\parallel=\mathrm{const}~,
\label{billII}
\eea
and
\bea
K^i\partial_i A=0~,~~~\tilde{\nabla}_{( i} K_{j )}=0~,
\label{ooo}
\eea
ie the length of the Killing spinors is constant, the warp factor is invariant and $\tilde K$ is Killing on the internal space $M^{d-k}$, where $\tilde \nabla$ is the Levi-Civita connection
of the internal space M$^{d-k}$.  To summarize, the conditions for $\mathfrak{g}_0$ to decompose as  $\mathfrak{g}_0=\mathfrak{so}(k-1,2)\oplus \mathfrak{t}_0$
are (\ref{billI}), (\ref{billIII}),  (\ref{billII}) and (\ref{ooo}).

\subsection{Global conditions for the decomposition of $\mathfrak{g}_0$ }

The conditions (\ref{billI}), (\ref{billIII}) and   (\ref{billII}) we have found  on the Killing spinor bilinears in the previous section for  $\mathfrak{g}_0=\mathfrak{so}(k-1,2)\oplus \mathfrak{t}_0$ can also derived in an elegant way after  imposing that the internal space M$^{d-k}$  is compact without boundary and the fields are smooth. Indeed after
 setting $\Lambda=\sigma_++\tau_+$, one can demonstrate  using the KSEs (\ref{kseI}), (\ref{kseII}) that
\bea
\nabla_i \parallel \Lambda\parallel^2=-{2\over\ell} \langle \sigma_+, \Gamma_i \Gamma_z \tau_+\rangle~,
\label{xxy}
\eea
and
\bea
\nabla^2 \parallel \Lambda\parallel^2+ 2 \partial^i\log A\, \nabla_i \parallel \Lambda\parallel^2=0~.
\eea
Applying the maximum principle, one concludes that  $\parallel \Lambda\parallel$ is constant which gives in particular (\ref{billII}).  Then  (\ref{xxy})
implies  (\ref{billI}).  As in the previous section  (\ref{billIII}) can be derived from (\ref{kseII}) using  (\ref{billI}). This establishes the assertion.

\section{AdS$_3$ in D=11 and Type II Theories} \label{sec:m_theory_type_ii_backgrounds}

\subsection{$N=2$}

 AdS$_3$ backgrounds preserve an even number of supersymmetries. Therefore the minimal case is that for which  a background preserves exactly two supersymmetries.  The Killing spinors can be given
 in terms of either $\sigma_\pm$ or $\tau_\pm$ spinors. In the former case, we find that the   Killing spinors are
 \bea
 \epsilon_1 = \sigma_+~, \qquad \epsilon_2 = A \Gamma_{- z} \sigma_+ + 2 \ell^{-1} u \sigma_+~,
 \label{killh2}
\eea
where we have set  $\sigma_- = A \Gamma_{- z} \sigma_+$.  In terms of $\tau_\pm$ the Killing spinors are as in (\ref{taukill}).
Using the results of appendix C, we find that the bilinears are given as
\bea
K_{11}= \lambda^-~,~~~
 K_{12}= -\lambda^z - \ell^{-1} M^{+ -}~,~~~
 K_{22}= -2 \lambda^+ - 4 \ell^{-1} M^{z +}~,
 \label{kvf2}
\eea
where we have chosen the normalization $2\| \sigma_+ \| ^2 = 1$ in order to simplify  coefficients and $\lambda^-, \lambda^z, M^{+ -}, M^{z +}$ are isometries of AdS$_3$
 given in appendix B.  Note that all 1-form bilinears have non-vanishing
components only along the AdS$_3$ directions.

The direct computation of   spinorial Lie derivatives in appendix C reveals that
\begin{align*}
 \mathcal{L}_{K_{11}} \epsilon_1 &= 0~, & \mathcal{L}_{K_{12}} \epsilon_1 &= -\ell^{-1} \epsilon^1~, & \mathcal{L}_{K_{22}} \epsilon_1 &= -2 \ell^{-1} \epsilon_2~,
 \\
 \mathcal{L}_{K_{11}} \epsilon_2 &= 2 \ell^{-1} \epsilon_1~, & \mathcal{L}_{K_{12}} \epsilon_2 &= \ell^{-1} \epsilon_2~, & \mathcal{L}_{K_{22}} \epsilon_2 &= 0~.
\end{align*}
Using this and the definition of  the symmetry superalgebra in (\ref{super}), we find that the non-vanishing commutators are
\bea
&&\{Q_A, Q_B\}=V_{AB}~,~~~ [V_{AB}, Q_{C}]=-\ell^{-1} (\epsilon_{CA} Q_{B}+\epsilon_{CB} Q_{A})~,
\cr
&&[V_{AB}, V_{A'B'}]= \ell^{-1} (\epsilon_{AA'} V_{BB'}+ \epsilon_{BA'} V_{AB'}+ \epsilon_{AB'} V_{BA'}+ \epsilon_{BB'} V_{AA'})~,
\label{superh2}
\eea
where $\epsilon_{AB}$, $A,B=1,2$, is the Levi-Civita tensor with $\epsilon_{12}=1$.   As the Killing spinors have the same form in all  10- and 11-dimensional supergravity theories, the superlgebra of  $N=2$ AdS$_3$ backgrounds in all these theories is   (\ref{superh2}).
The Lie algebra of the three Killing vectors $K_{AB}$ is $\mathfrak{sp}(2)=\mathfrak{sl}(2,\bR)=\mathfrak{so}(1,2)$ and acts on the two supersymmetry generators with the fundamental representation. This KSA is isomorphic\footnote{We mostly follow the notation \cite{kac} for superalgebras. However as  different real forms of a superalgebra  appear in the analysis,
to distinguish between them we have replaced   $\mathfrak{osp}(n,m)$  with $\mathfrak{osp}(n\vert m)$.  If the signature is $(n-2, 2)$,  we write
$\mathfrak{osp}(n-2, 2\vert m)$. If this is not sufficient to specify the real form we also use $\mathfrak{g}^*$ to denote a different real form from $\mathfrak{g}$.  In all cases we have
investigated the real form is specified by the real form of $\mathfrak{g}_0$ which we give in all cases. }
to $\mathfrak{osp}(1\vert 2)$.

\subsection{$N=4$} \label{sec:m_iia_iib_ads3_extended}

There are three ways to construct the four Killing spinors of $N=4$ AdS$_3$ backgrounds in D=11 and type II theories. For the first two options, one  can  choose  the Killing spinors to depend on four linearly independent $\sigma_\pm$ spinors or  four  linearly independent $\tau_\pm$ spinors.  In the third option, one chooses the first two Killing spinors to depend on $\sigma_\pm$
and the remaining two on $\tau_\pm$.

\subsubsection{Left and right superalgebra}

Let us begin with the third  possibility  where the first two Killing spinors  of $N=4$ AdS$_3$ backgrounds are  expressed in terms of $\sigma$ spinors as in (\ref{killh2})
while the remaining two are  expressed in terms of $\tau$ spinors as
\bea
\epsilon_3=e^{-{z\over\ell}} \tau_+-\ell^{-1} r A^{-1} e^{-{z\over\ell}}\Gamma_{-z} \tau_+~,~~~\epsilon_4= A e^{{z\over\ell}}\Gamma_{-z} \tau_+~,
\label{taukill}
\eea
where we have used that if $\tau_+$ is a Killing spinor then $\tau_-=A\Gamma_{-z} \tau_+$ is also a Killing spinor, and (\ref{ksp2}).

To find the superalgebra in this case, first note that the 1-form bilinears of the $\tau$ type Killing spinors are
\bea
 K_{33} = \lambda^- + 2 \ell^{-1} M^{z -}~,~~~
K_{34} = -\lambda^z + \ell^{-1} M^{+ -}~,~~~K_{44} = -2 \lambda^+~,
\label{tauiso}
\eea
where $\lambda^+, M^{z -}$ are also isometries of AdS$_3$, see appendix B. So  all Killing spinor bilinears lie along the AdS$_3$ subspace directions.

It remains to compute the rest of the 1-form bi-linears.  Using the orthogonality of $\sigma_+$ and $\tau_+$ spinors (\ref{billIII}) and (\ref{billI}) as well as the expressions
for the bilinears in appendix C, one can show that all the remaining bilinears vanish. Then a consequence of the super-Jacobi identity of the  superalgebra is that all the commutators between Killing spinors constructed from $\sigma$ spinors and their bilinears and those constructed from $\tau$ spinors and their bilinears vanish.   As a result, the KSA is $\mathfrak{g}=\mathfrak{g}_L\oplus \mathfrak{g}_R=\mathfrak{osp}(1\vert 2) \oplus \mathfrak{osp}(1\vert 2)$.
Viewing AdS$_3$  locally as a group manifold, $\mathfrak{g}_L$ is associated with the left action   while $\mathfrak{g}_R$ is associated with the right action on AdS$_3$.

\subsubsection{Left or right superalgebra}

Next suppose that all Killing spinors are expressed in terms of four linearly independent $\sigma_\pm$ spinors.  In this case,
the Killing spinors of AdS$_3$ backgrounds with extended supersymmetry are multiple copies of the Killing spinors (\ref{killh2}) that appear for the solutions
preserving two supersymmetries.  Because of this, it is convenient to denote the Killing spinors with a double index as $\epsilon_{Ar}$ where $A=1,2$  labels
the two spinors in the same multiplet and $r=1,\dots, N/2$ denotes the number of multiplets. Using this notation,
the Killing spinors of  AdS$_3$ backgrounds that preserve four supersymmetries ($N=4$) can be written as
\bea
 \epsilon_{1r} = \sigma_+^r~, ~~~ \epsilon_{2r} = A \Gamma_{- z} \sigma_+^r + 2 \ell^{-1} u \sigma_+^r ,
\label{hks2}
\eea
where $r=1,2$. We can assume without loss of generality that $\sigma_+^1$ and $\sigma_+^2$ are orthogonal.  From construction, they have to be linearly independent.  As their lengths and inner products  are
 constant and the KSE are linear over the real numbers, they can always be chosen as orthogonal via a Gram-Schmidt process.

An inspection of the results of appendix C illustrates that the 1-form bilinears of the above Killing spinors  can be written as
\bea
K_{Ar, Bs}=K_{AB} \delta_{rs}+ \epsilon_{AB} \tilde K_{rs}~,
\label{hbi2}
\eea
where $K_{AB}$ are as in (\ref{kvf2}) and $\tilde K_{rs}=-\tilde K_{sr}=\epsilon_{rs} \tilde K$ is a new Killing vector which has non-vanishing components only along the internal space M$^{d-3}$ directions
and depends only on the coordinates of the internal space.
After  choosing $2 \langle \sigma_+^r, \sigma_+^s\rangle=  \delta_{rs}$,  the anti-commutator of the odd
generators can be written as in
\bea
\{Q_{Ar}, Q_{Bs}\}=V_{AB} \delta_{rs}+ \epsilon_{rs} \epsilon_{AB}\tilde V~,
\eea
where we have set $\tilde V_{rs}=\epsilon_{rs} \tilde V$.

 All the commutators $[V_{AB}, Q_{Cr}]$ can be read from the results of appendix C.  It remains to
determine the commutator  $[\tilde V, Q_{Ar}]$.  As we do not have additional information on the geometry of the internal space M$^{d-3}$, this commutator cannot
be computed explicitly.  Instead, we shall utilize the closure of the KSA.
For this first observe that $\tilde V=\{Q_{11}, Q_{22}\}=-\{Q_{12}, Q_{21}\}$.  Thus for every choice of $Q_{Ar}$ there is another odd generator $Q_{A'r'}$ with $A\not=A'$ and $r\not= r'$ such that $\{Q_{Ar}, Q_{A'r'}\}\varpropto \tilde V$. Then the super-Jacobi identities imply that
\bea
[\tilde V,  Q_{Ar}]&\varpropto& [\{Q_{Ar}, Q_{A'r'}\}, Q_{Ar}]=-{1\over2} [\{Q_{Ar}, Q_{Ar}\}, Q_{A'r'}]=-{1\over2}[V_{AA}, Q_{A'r'}]
\cr &=&\ell^{-1} \epsilon_{A'A} Q_{Ar'}~.
\label{nneib}
\eea
Using this and after a brief computation, one can verify that the superalgebra is
\bea
&&\{Q_{Ar}, Q_{Bs}\}=V_{AB} \delta_{rs}+ \epsilon_{rs} \epsilon_{AB}\tilde V~,
\cr
&& [V_{AB}, Q_{Cr}]=-\ell^{-1} (\epsilon_{CA} Q_{Br}+\epsilon_{CB} Q_{Ar})~,~~~
\cr
&&[\tilde V, Q_{Ar}]= -\ell^{-1} \epsilon_r{}^s Q_{As}~,
\label{superh4}
\eea
 which is isomorphic to $\mathfrak{osp}(2\vert 2)$. The analysis for the case in which  all the Killing spinors depend on $\tau_\pm$ spinors follows in the same way.

Note that as there is some freedom to choose
$\sigma_+^2$, it is not a priori obvious that there is a non-vanishing 1-form bilinear $\tilde K_{rs}$  associated to the generator $\tilde V_{rs}$. However if $\tilde K_{rs}$ is chosen to vanish
and so the superalgebra does not have a  $\tilde V_{rs}$ generator, the super-Jacobi identities of three $Q_{Ar}$ generators are not satisfied. Therefore consistency
of the KSA requires the presence of the $\tilde V_{rs}$ generator.

\subsection{Some structure theory}

\subsubsection{Structure constants of KSAs}

To proceed further for $N>4$ let us suppose that all Killing spinors $\epsilon_{Ar}$  are constructed from linearly independent $\sigma^r_\pm$ spinors.
Then as we shall explain below the (anti)-commutators of the superagebra can be written as
\bea
\{Q_{Ar}, Q_{Bs}\}&=&\delta_{rs} V_{AB}+ \epsilon_{AB} \tilde V_{rs}~,
\cr
[V_{AB}, Q_{Cr}]&=&-\ell^{-1} (\epsilon_{CA} Q_{Br}+\epsilon_{CB} Q_{Ar})
\cr
[V_{AB}, V_{A'B'}]&=& \ell^{-1} (\epsilon_{AA'} V_{BB'}+ \epsilon_{BA'} V_{AB'}+ \epsilon_{AB'} V_{BA'}+ \epsilon_{BB'} V_{AA'})
\cr
[\tilde V_{rs}, Q_{At}]&=&-\ell^{-1}(\delta_{tr} Q_{As}- \delta_{ts} Q_{Ar})+\ell^{-1}\alpha_{rst}{}^p Q_{Ap}
\label{3x}
\eea
where as it will be explained below  $\alpha$ is a constant 4-form.
The first anti-commutator follows from the results of appendix C where we have normalized the spinors as $2\langle\sigma_+^r, \sigma_+^s\rangle =\delta_{rs}$.
  The bosonic generators $V_{AB}$ are associated with the left AdS$_3$ isometries, and the $\tilde V_{rs}$ are associated with isometries of
the internal space.  However note that $\tilde V_{rs}$  are not necessarily   linearly independent. The second and third commutators follow from a direct computation presented in appendices B and  C as the dependence of the Killing spinors and $K_{AB}$ on AdS$_3$ coordinates is known. Then the fourth commutator can be restricted by requiring  consistency with the super-Jacobi identities.

To justify the $[\tilde V_{rs}, Q_{At}]$ commutator observe that  if either $t=r$ or $t=s$, the commutator follows from the results established
in the $N=4$ case. Now suppose that  $t\not=r,s$ and consider   ${\cal L}_{\tilde K_{rs}} \epsilon_{At}$.  As we are investigating backgrounds preserving strictly N supersymmetries there must be constants $\alpha$ and $\tilde \alpha$ such that
\bea
{\cal L}_{\tilde K_{rs}} \epsilon_{At}= \alpha_{rst}{}^\ell  \epsilon_{A\ell}+ \epsilon_A{}^B  \tilde\alpha_{rst}{}^\ell  \epsilon_{B\ell}~,~~~t\not=r,s~,
\label{alpha6}
\ee
where $\alpha_{rst}{}^\ell=-\alpha_{srt}{}^\ell$ and $\tilde\alpha_{rst}{}^\ell=-\tilde\alpha_{srt}{}^\ell$.
As $\tilde K_{rs}$ are along the internal manifold, the spinorial Lie derivative preserves the dependence of the Killing spinors on the AdS$_3$ coordinates.  Therefore $\tilde\alpha=0$.
Furthermore, the super-Jacobi identity of the generators $Q_{A r}$, $Q_{Bs}$ and $Q_{A t}$, for $A\not= B$,  implies that $\alpha_{rst}{}^\ell=\alpha_{str}{}^\ell$. This
together with $\alpha_{rst}{}^\ell=-\alpha_{srt}{}^\ell$ gives
\bea
\alpha_{rst}{}^\ell=\alpha_{[rst]}{}^\ell~.
\label{con16x}
\eea
Furthermore as $\epsilon_{1t}=\sigma_+^t$ and ${\cal L}_{\tilde K_{rs}}\langle \sigma_+^t, \sigma_+^\ell\rangle=0$ , we find that
\bea
\alpha_{rst\ell}=-\alpha_{rs\ell t}~,
\label{con16xx}
\eea
where we have lowered the index with $\delta_{rs}$. Combining (\ref{con16x}) and (\ref{con16xx}), we deduce
that
\bea
\alpha_{rst\ell}=\alpha_{[rst\ell]}~,
\label{con16}
\eea
and so $\alpha$ is a 4-form.

The identification of KSAs of AdS$_3$ backgrounds for $N>4$ depends crucially on determining the form $\alpha$. One way to do this is to observe that
the outer automorphisms of  KSA  include the action of $SO(N/2)$ on  $Q_{Ar}$. With this action $\alpha$ transforms
as a 4-form.  As a result it suffices to consider representatives of the orbits of $\mathfrak{so}(N/2)$ on the space of 4-forms.  This consideration is sufficient
to identify all the KSAs for $N\leq 12$ and it is explored in appendix D.  However to find all KSAs, we investigate  the  structure of these KSAs further.

\subsubsection{The AdS$_3$ KSAs are direct sums of Left and Right KSAs}

Before we proceed further with the investigation of the KSA, let us consider the case where some of the Killing spinors are constructed from $\sigma$ spinors and  some others  from $\tau$ spinors.
A straightforward application of the computation presented in appendix C reveals that the superalgebra $\mathfrak{g}(\sigma)$  associated to the $\sigma$ type of Killing spinors and the
superalgebra $\mathfrak{g}(\tau)$ associated to the $\tau$ type of Killing spinors commute, and so we can set $\mathfrak{g}_L=  \mathfrak{g}(\sigma)$ and $\mathfrak{g}_R=  \mathfrak{g}(\tau )$; $\mathfrak{g}=\mathfrak{g}_L\oplus
\mathfrak{g}_R$. To see this observe that all mixed $\sigma$ and $\tau$ 1-form bilinears vanish. This implies that  the odd $\sigma$ type generators anti-commute with
the odd $\tau$ type generators. Furthermore the commutator of $\sigma$ type even generators associated to isometries on AdS$_3$ with $\tau$ type odd generators vanishes, and vice versa.
This can be seen from the spinorial Lie derivatives in appendix C.

It remains to demonstrate that the commutators of $\sigma$ ($\tau$) type even generators associated to isometries on the internal space with
$\tau$ ($\sigma$) type odd generators vanish as well.  First observe that Killing vectors along the internal space preserve the functional
dependence of Killing spinors on the AdS$_3$ coordinates. As $\sigma$ and $\tau$ Killing spinors have different such dependence it follows that they cannot be rotated to each
other under such spinorial Lie derivatives. Then  upon using  super-Jacobi identities and the fact that all  mixed bilinears vanish,  one can show that the commutator of
a $\sigma$ ($\tau$) internal even generator with any $\tau$ ($\sigma$) odd generator vanishes. This  establishes the result.

 A consequence of $\mathfrak{g}=\mathfrak{g}_L\oplus
\mathfrak{g}_R$ is that it suffices
to investigate the KSAs associated with only $\sigma$ type Killing spinors. Then $\mathfrak{g}$  can be easily found as $\mathfrak{g}_L$ and
$\mathfrak{g}_R$ are isomorphic.

\subsubsection{Structure theorems}

Let $\mathfrak{g}=\mathfrak{g}_L$ be the super-algebra of AdS$_3$ backgrounds preserving $N$ supersymmetries.
Decompose $\mathfrak{g}_0=\mathfrak{sp}(2)\oplus \mathfrak{t}_0$, where $\mathfrak{t}_0=\mathrm{Span} (\tilde V_{rs})$.
It is clear that $\mathfrak{g}_1=\bR^2\otimes \bR^{{N\over2}}$ and the action of $\mathfrak{t}_0$ preserves the Euclidean inner product on $ \bR^{{N\over2}}$.
As a result  $\mathfrak{t}_0\subseteq \mathfrak{so}(N/2)$.  We shall show that $\mathfrak{t}_0$ is associated with a subgroup of $SO(N/2)$ which acts transitively on the $S^{{N\over2}-1}$  sphere in $\bR^{{N\over2}}$.
To demonstrate this we shall first show the following.
\vskip 0.4cm

{\bf Proposition:} Given $u,w\in \bR^{{N\over2}}$ and $u,w$ linearly independent, then $\tilde V_{u\times w}\equiv  u^r w^s \tilde V_{rs}$ cannot vanish.
\vskip 0.2cm
{\bf Proof:}  Suppose that $\tilde V_{u\times w}=0$. In such a case it follows from (\ref{3x}) that
\bea
\{ u\cdot Q_A, w\cdot Q_B\}= u\cdot w~~V_{AB}
\eea
where $u\cdot w$ is the Euclidean inner product in $ \bR^{{N\over2}}$ and $ u\cdot Q_A=u^r Q_{Ar}$. Then upon using the super-Jacobi identity
\bea
&&u^2 \ell^{-1} \epsilon_{AB}  w\cdot Q_B= u^2 [V_{AB}, w\cdot Q_B]=[ \{u\cdot Q_A, u\cdot Q_B\},  w\cdot Q_B]
\cr
&&=-[ \{w\cdot Q_B, u\cdot Q_A\} , u\cdot Q_B]-[ \{u\cdot Q_B , w\cdot Q_B \} , u\cdot Q_A]
\cr
&&=
-u\cdot w ( [V_{AB},  u\cdot Q_B]+ [V_{BB}, u\cdot Q_A])=\ell^{-1} \epsilon_{AB} u\cdot w ~u\cdot Q_B
\eea
which is satisfied iff $u$ and $w$ are linearly dependent as $\mathrm{dim} \mathfrak{g}_1=N$. This a contradiction and so  $\tilde V_{u\times w}\not=0$.
\hfill{$\triangle$}
 \vskip 0.4cm

 {\bf Proposition:} The Lie algebra $\mathfrak{t}_0$ is associated with a subgroup $H_0$  of $SO(N/2)$ that acts transitively on the sphere $S^{{N\over2}-1}\subset \bR^{{N\over2}}$.

 \vskip 0.2cm
{\bf Proof:}
It suffices to show that given  two linearly independent vectors  $u,w\in \bR^{{N\over2}}$, there is element $R(u,w)\in \mathfrak{t}_0$ such that $R(u,w)$ generates $SO(2)$ rotations
on the 2-plane spanned by $u$ and $w$ in $\bR^{{N\over2}}$. As the  $SO(2)$ rotations act transitively on all directions in the 2-plane spanned by $u$ and $w$,
it follows that there is an element in $H_0$ which rotates the direction defined by the vector $u$ onto that of the vector $w$.

For this set $R(u,w)=\tilde V_{u\times w}$  and  observe that
\bea
[\tilde V_{u\times w}, p\cdot Q_A]=-\ell^{-1} ( p\cdot u~ w\cdot Q_A- p\cdot w~ u\cdot Q_A)~,
\eea
for any $p$ that lies in the 2-plane spanned by $u$ and $w$.
So indeed  $\tilde V_{u\times w}$ acts as an infinitesimal orthogonal rotation on the 2-plane spanned by $u$ and $w$.
As this can be done for any
$u,w\in \bR^{{N\over2}}$, it follows that $H_0$ acts transitively on $S^{{N\over2}-1}\subset \bR^{{N\over2}}$.

\hfill{$\triangle$}

\vskip 0.4cm

 {\bf Proposition:} The representation of  Lie algebra $\mathfrak{t}_0$ on $\mathfrak{g}_1$ leaves invariant the 4-form $\alpha$.
 \vskip 0.2cm
{\bf Proof:} For this write
\bea
[\tilde V_{rs}, Q_{At}]= D(\tilde V_{rs})_t{}^\ell Q_{A\ell}~.
\eea
It suffices to show that
\bea
D(\tilde V_{rs})_{[t_1}{}^\ell \alpha_{t_2t_3t_4]\ell}=0 \ .
\label{inv4form}
\eea
First using the super-Jacobi identities, one can establish that
\bea
[\tilde V_{rs}, \tilde V_{r's'}]&=&- \ell^{-1} \big( \delta_{rr'} \tilde V_{ss'}- \delta_{sr'}\tilde V_{rs'}-\delta_{rs'} \tilde V_{sr'}+\delta_{ss'} \tilde V_{rr'}
\cr
&&~~~~~-\alpha_{rsr'}{}^t \tilde V_{ts'}+
\alpha_{rss'}{}^t \tilde V_{tr'}\big)~.
\eea
As this bracket is skew-symmetric in the interchange of the pair $\tilde V_{rs}$ and $\tilde V_{r's'}$, one obtains the identity
\bea
\alpha_{rsr'}{}^t \tilde V_{ts'}-
\alpha_{rss'}{}^t \tilde V_{tr'}+\alpha_{r's'r}{}^t \tilde V_{ts}-
\alpha_{r's's}{}^t \tilde V_{tr}=0~.
\eea
Taking the commutator with $Q_{Ap}$, one arrives at (\ref{inv4form}).
\hfill{$\triangle$}
\vskip0.2cm

The results we have obtained  above can be summarized as follows.

\vskip 0.1cm

 {\bf Theorem:} The necessary conditions for a superalgebra $\mathfrak{g}$ to be the  KSA of AdS$_3$ backgrounds are  that  $\mathfrak{g}_0=\mathfrak{sp}(2)\oplus \mathfrak{t}_0$ and
 that $\mathfrak{g}_1=\bR^2\otimes \bR^{{N\over2}}$, where $\mathfrak{t}_0$ is  the Lie algebra of a group acting transitively on $S^{{N\over2}-1}\subset \bR^{{N\over2}}$.
 Furthermore,  the representation of the Lie algebra $\mathfrak{t}_0$ in $\mathfrak{g}_1$ leaves  the 4-form $\alpha$ invariant.
 \hfill{$\triangle$}
 \vskip 0.2cm

The groups that act effectively and transitively on spheres have been classified in \cite{ms} and have been listed in  table 1. This classification
also specifies the representation of the group that acts transitively on the vector space $\bR^n$ in which  $S^{n-1}$ is embedded. This is essential
for finding the AdS$_3$ KSAs as we shall explain below.

\begin{table}[h]
\begin{center}
\vskip 0.3cm
\underline {Lie Algebras of Groups Acting Transitively on Spheres}
 \vskip 0.3cm
 \begin{tabular}{|c|c|c|}
  \hline
  Algebra & Sphere & $N/2$
  \\ \hline
  $\mathfrak{so}(n)$  &$S^{n-1}$ & $n$
  \\ \hline
 $ \mathfrak{u}(n) $ & $S^{2n-1}$ & $2n$
  \\ \hline
 $ \mathfrak{su}(n)$  & $S^{2n-1}$ & $2n$
  \\ \hline
 $ \mathfrak{sp}^*(n)\oplus \mathfrak{sp}^*(1) $ & $S^{4n-1}$ & $4n$
  \\ \hline
  $ \mathfrak{sp}^*(n)\oplus \mathfrak{u}(1)$  & $S^{4n-1}$ &  $4n$
  \\ \hline
  $ \mathfrak{sp}^*(n)$  & $S^{4n-1}$ &  $4n$
  \\ \hline
  $\mathfrak{g}_2 $ & $S^6$ & $7$
  \\ \hline
 $ \mathfrak{spin}(7)$  & $S^7$ & $8$
  \\ \hline
 $ \mathfrak{spin}(9) $ & $S^{15}$ & $16$
  \\ \hline
 \end{tabular}
 \vskip 0.2cm
  \caption{  $ \mathfrak{sp}^*( n)$ is the compact symplectic algebra with (real) dimension $n (2n+1)$ and a real form of $ \mathfrak{sp}(2 n)$.}
 \end{center}
\end{table}

To identify $\mathfrak{t}_0$ with the Lie algebras of the groups listed in table 1, it remains to find the conditions for $\mathfrak{t}_0$ to act effectively on $\mathfrak{g}_1$. For this
 define the subalgebra $\mathfrak{c}\subset \mathfrak{t}_0$ such that
$[\mathfrak{c}, \mathfrak{g}_1]=0 $ or equivalently
\bea
 \mathfrak{c}=\{u^{rs} \tilde V_{rs}\in \mathfrak{t}_0~~\vert ~~2u^{rs}- u^{pq} \alpha_{pq}{}^{rs}=0\}~.
 \eea
Observe that $\mathfrak{c}$ is a commutative ideal\footnote{Super-Jacobi identities put additional restrictions on  $\mathfrak{c}$ which may lead to the vanishing of all its elements.}   of $\mathfrak{g}$, $\mathfrak{g}_0/\mathfrak{c}=\mathfrak{sp}(2)\oplus \mathfrak{t}_0/\mathfrak{c}$ and $\mathfrak{t}_0/\mathfrak{c}$ acts effectively
 on $\mathfrak{g}_1$. As $\mathfrak{t}_0/\mathfrak{c}$ also acts transitively on the spheres, it can be identified with the Lie algebra of the groups listed in table 1.
 Significantly, the representation of $\mathfrak{t}_0/\mathfrak{c}$ on $\mathfrak{g}_1$ is determined from that of the groups listed in table 1 on $\bR^{{N\over2}}$ in which the sphere
 $S^{{N\over2}-1}$ is embedded.  If $\mathfrak{c}$ is non-empty, then  $\mathfrak{g}$ is not simple.  This is the reason that non-simple superalgebras can occur
 as KSAs for AdS$_3$ backgrounds\footnote{To our knowledge there does not exist an example of an AdS$_3$ solution with a non-simple KSA.  So such solutions may not exist.  However,
 they cannot be ruled out within our framework as their exclusion  requires a more detailed description of the geometry of the internal space.}

It should be noted that the KSA of AdS$_3$ backgrounds admits a consistent, supersymmetric and invariant inner product\footnote{For the definition see \cite{kac}.} given by
\bea
&&\langle Q_{Ar}, Q_{Bs}\rangle=\epsilon_{AB} \delta_{rs}~,~~~\langle V_{AB},V_{A'B'}\rangle=-\ell^{-1}\epsilon_{AA'} \epsilon_{BB'}-\ell^{-1}\epsilon_{BA'} \epsilon_{AB'}~,
\cr
&&\langle \tilde V_{rs}, \tilde V_{r's'}\rangle=\ell^{-1}[\delta_{rr'} \delta_{ss'}-\delta_{sr'} \delta_{rs'}-\alpha_{rsr's'}]~.
\label{innerpro}
\eea
Observe that if $\mathfrak{c}\not=\emptyset$, this inner product is degenerate.

\subsection{KSAs from $\mathfrak{so}(n)$ acting transitively on $S^{n-1}$ }

\subsubsection{Generic case $N=N_\sigma=2n$}

 If  $N\not= 8$, then $\mathfrak{so}(n)$ does not admit an invariant 4-form in the fundamental $n$-dimensional representation
and thus $\alpha=0$. Then
it is  straightforward to observe that the KSA  is isomorphic to $\mathfrak{osp}(n\vert 2)$.  The same applies in the $N=8$ case provided we choose $\alpha=0$.
As there are no maximally supersymmetric AdS$_3$ backgrounds $n<16$.  The algebra of isometries of the internal space is $\mathfrak{so}(n)$.

\subsubsection{$N=N_\sigma=8$}

  As $\alpha$ is a 4-form (\ref{alpha6}) in a 4-dimensional space it is proportional to the volume form $\epsilon$. So we write
\bea
\alpha_{rst\ell}=\hat\alpha\, \epsilon_{rst\ell}~,~~~\
\eea
for some constant $\hat\alpha$.
Using this, the non-vanishing
(anti-)commutators (\ref{3x}) of KSA are
\bea
\{Q_{Ar}, Q_{Bs}\}&=&V_{AB} \delta_{rs}+ \epsilon_{AB}\tilde V_{rs}~,
\cr
 [V_{AB}, Q_{Cr}]&=&-\ell^{-1} (\epsilon_{CA} Q_{Br}+\epsilon_{CB} Q_{Ar})~,~~~
 \cr
[\tilde V_{rs}, Q_{At}]&=&-\ell^{-1}(\delta_{tr} Q_{As}- \delta_{ts} Q_{Ar})+\ell^{-1}\hat\alpha\, \epsilon_{rst}{}^p Q_{Ap}
\cr
[\tilde V_{rs}, \tilde V_{r's'}]&=&- \ell^{-1} \big( \delta_{rr'} \tilde V_{ss'}- \delta_{sr'}\tilde V_{rs'}-\delta_{rs'} \tilde V_{sr'}+\delta_{ss'} \tilde V_{rr'}
\cr
&&~~~~~-\hat\alpha \epsilon_{rsr'}{}^t \tilde V_{ts'}+
\hat\alpha \epsilon_{rss'}{}^t \tilde V_{tr'}\big)~,
\label{d2acom}
\eea
where we have neglected the commutators of the $V_{AB}$ already given in (\ref{superh2}).
This algebra is isomorphic to a real form of the $\mathfrak{D}(2,1;\alpha)$ superalgebra with  ${\alpha} = \frac{\hat\alpha+1}{1-\hat\alpha}$ and $\hat\alpha\not=0,1,-1$.  The isometry algebra of the internal space is
$\mathfrak{so}(3)\oplus \mathfrak{so}(3)$.
If $\hat{\alpha} = 0$, this superalgebra is isomorphic to  $\mathfrak{osp}(4\vert 2)$ as expected.

It remains to investigate the cases for which   $\hat{\alpha}$  is either $1$ or $-1$. For this write $\tilde V_{rs}=\tilde V^+_P (\omega_P^{(+)})_{rs}+ \tilde V_P^- (\omega_P^{(-)})_{rs}$ where $\omega^{(\pm)}$ are orthonormal  bases in the space  of self-dual
and anti-self dual two forms $\omega_{Prs}^{(\pm)} \omega_{Q}^{(\pm)rs}=4 \delta_{PQ}$. Then $[\tilde V_{rs}, Q_{At}]$ can be rewritten as
\bea
[\tilde V^+_P, Q_{At}]&=&{1\over2}\ell^{-1} (\hat\alpha-1) (\omega^{(+)}_P)_t{}^s Q_{As}
\cr
[\tilde V^-_P, Q_{At}]&=&-{1\over2}\ell^{-1} (\hat\alpha+1) (\omega^{(-)}_P)_t{}^s Q_{As} \ .
\label{d21aa}
\eea
 If either $\tilde V^+$  or $\tilde V^-$ vanishes
 for either $\hat\alpha=1$ or $\hat\alpha=-1$, respectively, the superalgebra is a real form of $\mathfrak{sl}(2\vert 2)/1_{4\times 4}$.  The isometry algebra of the internal space is  $\mathfrak{so}(3)$.

 Next consider the possibility that both  $\tilde V^+$  and  $\tilde V^-$ do not vanish and either $\hat\alpha=1$ or $\hat\alpha=-1$. In such a case,  it can be seen that $\tilde V^+$ for  $\hat\alpha=1$ or $\tilde V^-$ for $\hat\alpha=-1$ become central and they are allowed to be non-vanishing as they do not appear
 on the right-hand-side of  $\mathfrak{so}(3)$ commutators generated by either $\tilde V^-$ or $\tilde V^+$, respectively.
We shall denote the resulting superalgebra with $\mathfrak{csl}(2\vert 2; 3)/1_{4\times4}$ where the  last numerical entry denotes the maximal number
of central generators.  This is not a (semi-)simple superalgebra.
It is not apparent that such a superalgebra  arises as a possibility in actual solutions.  However it cannot be ruled out on the grounds of symmetry and the geometric assumptions we have made.

\subsection{KSAs from  $\mathfrak{u}(n)$ acting transitively on $S^{2n-1}$}

In this case  $N=N_\sigma=4n$. Define an embedding of $\mathfrak{u}(n)$ into $\mathfrak{so}(2n)$ by choosing a complex structure $I$ in $\bR^{2n}$ compatible with the Euclidean metric and
with associated Hermitian form $\omega$, $\omega_{rs}=\delta_{rt} I^t{}_s$.  The 4-form $\alpha$ can be chosen as
\bea
\alpha=\hat\alpha \,\,\omega\wedge \omega~,
\eea
where $\hat\alpha$ is a constant.  The subalgebra $\mathfrak{t}_0$ decomposes as
\bea
\mathfrak{t}_0=\mathfrak{t}_0^{(1,1)}\oplus \mathfrak{t}_0^{(2,0)+(0,2)}
\eea
where $\mathfrak{t}_0^{(2,0)+(0,2)}$ and  $\mathfrak{t}_0^{(1,1)}$ are the
spaces of (2,0)- and (0,2)-forms  and   (1,1)-forms  with respect to $I$, respectively.   So $\mathfrak{t}_0^{(1,1)}= \mathfrak{u}(n)$.
The projectors are given by
\bea
(P^{(1,1)})^{r's'}_{rs}={1\over2} (\delta^{r'}{}_{[r} \delta^{s'}{}_{s]}+I^{r'}{}_{[r} I^{s'}{}_{s]})~,~~~(P^{(2,0)+(0,2)})^{r's'}_{rs}={1\over2} (\delta^{r'}{}_{[r} \delta^{s'}{}_{s]}-I^{r'}{}_{[r} I^{s'}{}_{s]}) \ .
\label{proI}
\eea
If both $\tilde V^{(1,1)}$ and $\tilde V^{(2,0)+(0,2)}$ act effectively on the $Q$'s, then consistency requires that $\mathfrak{so}(2n)$ acts with
the fundamental representation on the $Q$'s and the KSA must be isomorphic to $\mathfrak{osp}(2n\vert 2)$. Alternatively only the subalgebra $\mathfrak{u}(n)$ acts effectively
on the $Q$'s.  Imposing this by requiring that the elements of  $\mathfrak{t}_0^{(2,0)+(0,2)}$ commute with the $Q$'s gives that
\bea
\hat\alpha={1\over2}~.
\eea
Furthermore one finds that
\bea
[\tilde V^{(1,1)}_{rs}, Q_{At}]=-\ell^{-1} (\delta_{t r} Q_{As}-\delta_{t s} Q_{Ar}+ \omega_{tr} I^p{}_s  Q_{Ap}-  \omega_{ts} I^p{}_r  Q_{Ap}+\omega_{rs} I^p{}_t  Q_{Ap})~.
\label{sunrep}
\eea
Observe that although (\ref{sunrep}) does not give
the standard action of $\mathfrak{u}(n)$ on $\bC^n$ because of the last term in the commutator, this can be achieved after the change of basis
\bea
\tilde V^{(1,1)}_{rs}\rightarrow \tilde V^{(1,1)}_{rs}+{1\over 2(2-n)}\omega_{rs} \omega^{pq} \tilde V^{(1,1)}_{pq}~, ~~~n\not=2~.
\eea
Factoring with the ideal generated by $\mathfrak{c}=\tilde V^{(2,0)+(0,2)}$, the resulting KSA is $\mathfrak{sl}(n\vert 2)$.

A consequence of (\ref{sunrep}) is that
\bea
[\omega^{rs} \tilde V^{(1,1)}_{rs}, Q_{At}]=2\ell^{-1} (2-n) I^p{}_t  Q_{Ap}~,
\label{utrace}
\eea
ie the $\omega$-trace of the $\tilde V^{(1,1)}$ has a non-trivial action on the $Q$'s for $n\not=2$.  As a result for $n\not=2$, there is no a KSA
which is associated with $\mathfrak{su}(n)$ Lie algebra\footnote{There is an exception to this for $n=4$ where additional invariant 4-forms exist associated
 with the holomorphic (4,0)-form.  Consideration of these leads to the $\mathfrak{spin}(7)$ case that we shall investigate below.} in table 1 as this would require that $\omega^{rs} \tilde V^{(1,1)}_{rs}$ commutes with the $Q$'s.
However for $n=2$, one can further factor with
$\omega^{pq} \tilde V^{(1,1)}_{pq}$ yielding the superalgebra $\mathfrak{sl}(2\vert 2)/1_{4\times 4}$.

\subsection{KSAs from $\mathfrak{g}_2$ and $\mathfrak{spin}(7)$}

The argument required to identify the KSA in these two cases proceeds as in the $\mathfrak{u}(n)$ case. In the $\mathfrak{g}_2\subset \mathfrak{so}(7)$ case for which $N=N_\sigma=14$, the invariant 4-form
can be chosen as $\alpha=\hat\alpha\, \phi$ where $\hat\alpha$ is a constant and $\phi$ is the fundamental invariant 4-form of $\mathfrak{g}_2$. Moreover  $\mathfrak{t}_0=\mathfrak{t}^{\bf 14}_0\oplus \mathfrak{t}^{\bf 7}_0$, where  $\mathfrak{t}^{\bf 14}_0=\mathfrak{g}_2$.  The projectors are
\bea
(P_{{\bf 14}})^{r's'}_{rs}={2\over3} (\delta^{r'}{}_{[r} \delta^{s'}{}_{s]}+{1\over4} \phi_{rs}{}^{r's'})~,~~~(P_{{\bf 7}})^{r's'}_{rs}= {1\over3}(\delta^{r'}{}_{[r} \delta^{s'}{}_{s]}-{1\over2} \phi_{rs}{}^{r's'})~,
\eea
where $r,s, r', s'=1,\dots, 7$.
It is clear from this  that  $P_{{\bf 7}}\tilde V\in \mathfrak{t}^{\bf 7}_0$ commute with the $Q$'s provided that $\hat\alpha=-1/2$. The  superalgebra $\mathfrak{g}/\mathfrak{t}^{\bf 7}_0$ is isomorphic  to  $\mathfrak{g}(3)$ and $\mathfrak{c}=\mathfrak{t}^{\bf 7}_0$.
 $P_{{\bf 7}} \tilde V$ must vanish because if they do not, these generators appear as central
  extensions of $\mathfrak{g}_2$ which is  simple and so it does not admit such an extension.

In the $\mathfrak{spin}(7)\subset \mathfrak{so}(8)$ case for which $N=N_\sigma=16$, the invariant 4-form is the fundamental, invariant, self-dual form $\psi$.  So we choose $\alpha=\hat\alpha\,\, \psi$.  Furthermore
 $\mathfrak{t}_0=\mathfrak{t}^{\bf 21}_0\oplus \mathfrak{t}^{\bf 7}_0$ with $\mathfrak{t}^{\bf 21}_0= \mathfrak{spin}(7)$.   The projectors are
\bea
(P_{{\bf 21}})^{r's'}_{rs}={1\over4} (3\delta^{r'}{}_{[r} \delta^{s'}{}_{s]}+{1\over2} \psi_{rs}{}^{r's'})~,~~~(P_{{\bf 7}})^{r's'}_{rs}= {1\over4}(\delta^{r'}{}_{[r} \delta^{s'}{}_{s]}-{1\over2} \psi_{rs}{}^{r's'})~,
\eea
where now $r,s, r', s'=1,\dots, 8$.  For  $P_{{\bf 7}}\tilde V\in\mathfrak{t}^{\bf 7}_0$   to commute with the $Q$'s, one has to set $\hat\alpha=-1/3$.  The  superalgebra $\mathfrak{g}/\mathfrak{t}^{\bf 7}_0$ is isomorphic  to  $\mathfrak{f}(4)$ and $\mathfrak{c}=\mathfrak{t}^{\bf 7}_0$.
The super-Jacobi identity of the even generators requires that one has to set $\mathfrak{t}^{\bf 7}_0$ to zero as $\mathfrak{spin}(7)$ does not admit  central extensions.


\subsection{KSAs from $\mathfrak{sp}^*(n)\oplus \mathfrak{sp}^*(1)$ and the remaining cases}

\subsubsection{$\mathfrak{spin}(9)$, $\mathfrak{sp}^*(n)$ and $\mathfrak{sp}^*(n)\oplus \mathfrak{u}(1)$}

The transitive action of  $\mathfrak{spin}(9)$ on $S^{15}$ does not give rise to a KSA for AdS$_3$ backgrounds as there are no AdS$_3$ backgrounds preserving 32 supersymmetries.  It remains to investigate the cases $\mathfrak{sp}^*(n)$, $\mathfrak{sp}^*(n)\oplus \mathfrak{u}(1)$ and $\mathfrak{sp}^*(n)\oplus \mathfrak{sp}^*(1)$ for which $N=N_\sigma=4n$.

 To begin, the embedding of $\mathfrak{sp}^*(n)$ in $\mathfrak{so}(4n)$ is specified by a hyper-complex structure $I,J$ and $K$, $I^2=J^2=-1$,  $IJ=-JI$ and $K=IJ$ in $\bR^{4n}$. The generators of $\mathfrak{sp}^*(n)$ are those of $\mathfrak{so}(4n)$ which are (1,1) with respect to all complex structures. Thus we write $\mathfrak{t}_0=\hat{\mathfrak{t}}_0^{(1,1)}\oplus \mathfrak{m}_0$,
 where $\hat{\mathfrak{t}}_0^{(1,1)}=\mathfrak{sp}^*(n)$ and $\mathfrak{m}_0$ is the orthogonal complement.

 The most general
4-form $\alpha$ which is invariant under  $\mathfrak{sp}^*(n)$  can be chosen as
\bea
\alpha= a_1\, I\wedge I+ a_2\, J\wedge J+a_3\, K\wedge K+b_1\,  I\wedge J+ b_2\,  I\wedge K+b_3\,  J\wedge K~,
\label{aaform}
\eea
where we denote with the same symbol the complex structures and their associated Hermitian 2-forms, and  $a_1, a_2, a_3, b_1, b_2, b_3$ are constants.  Next we impose the condition that
the elements of $\mathfrak{m}_0$ commute with the $Q$'s. In particular, we impose the condition that the elements of $\mathfrak{t}_0$ which are (2,0) and (0,2) with respect to $I$, and so lie in $\mathfrak{m}_0$, must commute with the $Q$'s. We find
\bea
&&2({1\over2}-a_1)\big( \delta_{p[r} \delta_{s]q}+ I_{p[r} I_{s]q}\big)- {1\over2} a_2 \big( (J\wedge J)_{rspq}-(K\wedge K)_{rspq}+2K_{rs} K_{pq}+ 2J_{rs} J_{pq}\big)
\cr
&&- {1\over2} a_3 \big( -(J\wedge J)_{rspq}+(K\wedge K)_{rspq}+2K_{rs} K_{pq}+2 J_{rs} J_{pq}\big)
\cr &&
-{1\over2} b_1 \big((I\wedge J)_{rspq}+J_{rs} I_{pq}-I_{rs} J_{pq}+ 2\delta_{q[r} K_{s]p}-2K_{q[r} \delta_{s]p}\big)
\cr &&
-{1\over2} b_2 \big((I\wedge K)_{rspq}+K_{rs} I_{pq}-I_{rs} K_{pq}- 2\delta_{q[r} J_{s]p}+2 J_{q[r} \delta_{s]p}\big)
\cr &&
- b_3 (J\wedge K)_{rspq}=0~.
\eea
Taking the trace with $I_{pq}$, one gets that $b_1=b_2=0$ for $n>1$. Taking the trace with $J_{pq}$ one finds that
\bea
{1\over2}-a_1-(2n-1) a_2-a_3=0~,~~~b_3=0 \ .
\eea
Next skew-symmetrizing in all $r,s,p$ and $q$ indices and  considering the $(4,0)$ and $(2,2)$ parts with respect to $I$, one deduces that after using the
above equation that
\bea
a_1={1\over2}~,~~~a_2=a_3=0 \ .
\eea
Thus if $n\not=1$, the conditions above imply $b_1=b_2=b_3=a_2=a_3=0$ and $a_1={1\over2}$. Thus, the commutator $[\tilde V, Q]$ is as in the $\mathfrak{u}(2n)$ case.  However   $\mathfrak{sp}^*(n)\subset \mathfrak{su}(2n)$ and so the trace with respect to $I$ must vanish as well. We have seen that this is not possible.  So there is no  KSA that can be constructed using the action of $\mathfrak{sp}^*(n)$ on the spheres.

The embedding of $\mathfrak{sp}^*(n)\oplus \mathfrak{u}(1)$ in $\mathfrak{so}(4n)$ is again characterized by the complex structures $I$, $J$ and $K$ but now
 $\tilde V_{rs}\in \mathfrak{sp}^*(n)\oplus \mathfrak{u}(1)$ iff  $\tilde V_{rs}$ is (1,1) with respect to $I$ and the I-traceless part of $\tilde V_{rs}$,  $\tilde V_{rs}-{1\over4n} I_{rs} I^{r's'} \tilde V_{r's'}$,
is (1,1) with respect to both $J$ and $K$.  The most general $\mathfrak{sp}^*(n)\oplus \mathfrak{u}(1)$ invariant 4-form is
\bea
\alpha=a_1 I\wedge I+ a_2 (J\wedge J+K\wedge K)~,
\eea
ie $a_2=a_3$ and $b_1=b_2=b_3=0$ in (\ref{aaform}).
As $\tilde V_{rs}$ is (1,1) with respect to $I$, one can repeat the computation above to show that $a_2=a_3=0$ and  $a_1={1\over2}$. To demonstrate that a superalgebra cannot be constructed
from the $\mathfrak{sp}^*(n)\oplus \mathfrak{u}(1)$ action on a sphere,   take a complex basis with respect to $I$ to find
\bea
[\tilde V_{\beta\bar\gamma}, Q_{A\zeta}]=-\ell^{-1} (-2 \delta_{\zeta\bar \gamma}  Q_{A\beta}+ \delta_{\beta\bar\gamma} Q_{A\zeta})~.
\eea
This allows the computation of the commutator of $\tilde V_{rs}-{1\over4n} I_{rs} I^{r's'} \tilde V_{r's'}$ on the $Q$'s.  Imposing next that the $(2,0)+(0,2)$ component of  $\tilde V_{rs}-{1\over4n} I_{rs} I^{r's'} \tilde V_{r's'}$ with respect to $J$ has to commute with the $Q$'s  leads to a contradiction.  There are no KSA associated to the $\mathfrak{sp}^*(n)\oplus \mathfrak{u}(1)$ case.

\subsubsection{$\mathfrak{sp}^*(n)\oplus \mathfrak{sp}^*(1)$}

The elements in  $\mathfrak{sp}^*(n)\oplus \mathfrak{sp}^*(1)\subseteq \mathfrak{t}_0$  can be written as
\bea
\tilde V_{rs}=\hat{ V}_{rs}+ W_I I_{rs}+W_J J_{rs}+W_K K_{rs}~,
\eea
where $\hat{ V}_{rs}\in \mathfrak{sp}^*(n) $ are
 (1,1) with respect to all $I$, $J$ and $K$  while $W_I$, $W_J$ and $W_K$ are the generators of $\mathfrak{sp}^*(1)$.   The invariant 4-form is
\bea
\alpha=a\, (I\wedge I+  J\wedge J+K\wedge K)~,
\eea
ie $a_1=a_2=a_3=a$ and $b_1=b_2=b_3=0$ in (\ref{aaform}).

To proceed, consider the decomposition $\mathfrak{t}_0=\mathfrak{sp}^*(n)\oplus \mathfrak{sp}^*(1)\oplus \mathfrak{m}_0$.  We   have to demonstrate that
the elements of
$\mathfrak{m}_0$ commute with the $Q$'s, i.e.$[\mathfrak{m}_0, \mathfrak{g}_1]=0$.  Following (\ref{proI}) denote the projections onto the $(2,0)+(0,2)$ and $(1,1)$ subspaces of the space of 2-forms with respect to the complex
 structure $I$ with $P^{(2,0)+(0,2)}_I$ and  $P^{(1,1)}_I$, respectively, and similarly for $J$ and $K$.  Then   note that
 \bea
 P^{(2,0)+(0,2)}_I \mathfrak{t}_0= \mathfrak{w}_J\oplus \mathfrak{w}_K\oplus  P^{(2,0)+(0,2)}_I \mathfrak{m}_0~,
 \label{proI1}
 \eea
 and
 \bea
 P^{(2,0)+(0,2)}_J \mathfrak{t}_0= \mathfrak{w}_I\oplus \mathfrak{w}_K\oplus  P^{(2,0)+(0,2)}_J \mathfrak{m}_0~,
 \label{proI2}
 \eea
 with
 \bea
 \mathfrak{m}_0= P^{(2,0)+(0,2)}_I \mathfrak{m}_0 + P^{(2,0)+(0,2)}_J \mathfrak{m}_0~,
 \label{proI3}
 \eea
  where $\mathfrak{w}_I$,  $\mathfrak{w}_J$ and  $\mathfrak{w}_K$ are the subspaces spanned by the generators $W_I$, $W_J$ and $W_K$, respectively, ie
  $\mathfrak{sp}^*(1)=\mathfrak{w}_I\oplus \mathfrak{w}_J\oplus \mathfrak{w}_K$.
  Using (\ref{proI1}) and (\ref{proI2}) and after setting $a=1/2$,  we find that
  \bea
  [(P^{(2,0)+(0,2)}_I \tilde V)_{rs}, Q_{At}]=\ell^{-1}( J_{rs} J_t{}^p Q_{Ap}+K_{rs} K_t{}^p Q_{Ap})~,
  \cr
   [(P^{(2,0)+(0,2)}_J \tilde V)_{rs}, Q_{At}]=\ell^{-1}( I_{rs} I_t{}^p Q_{Ap}+K_{rs} K_t{}^p Q_{Ap})~.
  \eea
  Therefore we deduce that
  \bea
  [W_J, Q_{At}]=\ell^{-1} J_t{}^p Q_{Ap}~,~~[W_K, Q_{At}]=\ell^{-1} K_t{}^p Q_{Ap}~,~~[W_I, Q_{At}]=\ell^{-1} I_t{}^p Q_{Ap}~,
  \label{wwwq}
  \eea
 and
  \bea
 [ P^{(2,0)+(0,2)}_I \mathfrak{m}_0, \mathfrak{g}_1]=0~,~~~ [ P^{(2,0)+(0,2)}_J \mathfrak{m}_0, \mathfrak{g}_1]=0~.
 \eea
Then it follows from (\ref{proI3})  that $[  \mathfrak{m}_0, \mathfrak{g}_1]=0$.

If the elements of $\mathfrak{m}_0$ are set to zero,  the resulting  superalgebra
 is isomorphic to the real form,  $\mathfrak{osp}^*(4\vert 2n)$,  of $\mathfrak{osp}(4\vert 2n)$.  To see this  note  that  $\mathfrak{osp}^*(4\vert 2n)_0=\mathfrak{sp}(2)\oplus\mathfrak{sp}^*(1)\oplus\mathfrak{sp}^*(n)$
while  $\mathfrak{osp}(4\vert 2n)_0=\mathfrak{sp}(2n)\oplus \mathfrak{so}(4)=\mathfrak{sp}(2n)\oplus \mathfrak{so}(3)\oplus  \mathfrak{so}(3)$. Then $\mathfrak{sp}^*(n)$ and $\mathfrak{sp}(2)$  are  real forms
of $\mathfrak{sp}(2n)$  and $\mathfrak{so}(3)$, respectively, and $\mathfrak{sp}^*(1)=\mathfrak{so}(3)$. On the other hand if the elements of $\mathfrak{m}_0$ do not vanish, the KSA may have central terms.
It is also known that there are no maximally supersymmetric AdS$_3$ backgrounds and so $n$ is restricted as $n< 4$.
This completes the identification of all KSA for AdS$_3$ backgrounds.  In appendix D, we present some examples for cases with a low number of supersymmetries without using the classification
results of \cite{ms} that  confirm the results we have presented.

We conclude this section by stating all superalgebras of AdS$_3$ backgrounds which preserve 16 supersymmetries.  After taking into account the possibility
of having both $\sigma$ and $\tau$ Killing spinors can occur, $\mathfrak{g}= \mathfrak{g}_L\oplus \mathfrak{g}_R$, and setting the central terms to zero, one finds that the KSAs are
\bea
&&\mathfrak{osp}(8\vert 2)~,~~~\mathfrak{sl}(4\vert 2)~,~~~\mathfrak{f}(4)~,~~~ \mathfrak{osp}^*(4\vert 4)~,~~~\mathfrak{osp}(1\vert 2)\oplus \mathfrak{g}(3)~,~~~\mathfrak{osp}(1\vert 2)\oplus \mathfrak{osp}(7\vert 2)~,
\cr
&&\mathfrak{osp}(2\vert 2)\oplus\mathfrak{osp}(6\vert 2)~,~~~\mathfrak{osp}(2\vert 2)\oplus\mathfrak{sl}(3\vert 2)~,~~~\mathfrak{osp}(3\vert 2)\oplus \mathfrak{osp}(5\vert 2)~,~~~
\cr
&&\mathfrak{osp}(4\vert 2)\oplus \mathfrak{osp}(4\vert 2)~,~~~\mathfrak{osp}(4\vert 2)\oplus \mathfrak{sl}(2\vert 2)/1_{4\times 4}~,~~~\mathfrak{sl}(2\vert 2)/1_{4\times 4}\oplus \mathfrak{sl}(2\vert 2)/1_{4\times 4}~,~~~
\cr
&&\mathfrak{osp}(4\vert 2)\oplus \mathfrak{D}(2,1, \alpha)~,~~~\mathfrak{sl}(2\vert 2)/1_{4\times 4}\oplus \mathfrak{D}(2,1, \alpha)~,
\mathfrak{D}(2,1, \alpha)\oplus \mathfrak{D}(2,1, \alpha)~,
\eea
where we have stated the unordered pairs.  Otherwise one has for example to include both $\mathfrak{osp}(1\vert 2)\oplus \mathfrak{g}(3)$ and $\mathfrak{g}(3) \oplus  \mathfrak{osp}(1\vert 2)$ as
distinct possibilities.
Similar lists can be obtained for any number of supersymmetries.

\begin{table}[h]
\begin{center}
\vskip 0.3cm
\underline {  AdS$_3$ KSAs in type II and $d=11$}
 \vskip 0.3cm
 \begin{tabular}{|c|c|c|}
  \hline
  $N_\sigma/2$ & $\mathfrak{g}_L/\mathfrak{c}$& $\mathfrak{t}_0/\mathfrak{c}$
  \\ \hline
  $n$  &$\mathfrak{osp}( n\vert 2)$ &$\mathfrak{so}( n)$
  \\ \hline
 $ 2n, n>1 $ & $\mathfrak{sl}(n\vert 2)$ & $\mathfrak{u}(n)$
  \\ \hline
 $4n, n>1$  &  $ \mathfrak{osp}^*(4\vert 2n)$ &  $ \mathfrak{sp}^*(n)\oplus \mathfrak{sp}^*(1)$
 \\ \hline
$8$  &  $ \mathfrak{f}(4)$ &  $ \mathfrak{spin}(7)$
  \\ \hline
  $7$  &  $\mathfrak{g}(3) $ & $\mathfrak{g}_2 $
  \\ \hline
  $4$  &  $ \mathfrak{D}(2,1,\alpha)$ &  $ \mathfrak{so}(3)\oplus \mathfrak{so}(3)$
  \\ \hline
  $4$  &  $ \mathfrak{sl}(2\vert 2)/1_{4\times 4}$ &  $ \mathfrak{so}(3)$
  \\ \hline
 \end{tabular}
 \vskip 0.2cm
  \caption{  In all cases, $(\mathfrak{g}_L/\mathfrak{c})_0=\mathfrak{so}(1,2)\oplus \mathfrak{t}_0/\mathfrak{c}$.  It is required that  $N_\sigma/2<16$ as there are
  no maximally supersymmetric AdS$_3$ backgrounds.}
 \end{center}
\end{table}

\section{Heterotic Backgrounds} \label{sec:heterotic_backgrounds}

\subsection{Killing spinors}

  Under some mild
assumptions, the heterotic string supergravity admits only AdS$_3$ solutions and  the warp factor is constant \cite{hetads}. The solutions preserve 2, 4, 6 and 8 supersymmetries. The Killing spinors of such backgrounds are either
expressed in terms of $\sigma_\pm$ or $\tau_\pm$ spinors as in (\ref{killh2}) or (\ref{taukill}), respectively.  As only either $\sigma_\pm$ or $\tau_\pm$ Killing spinors can occur,
we shall focus on the Killing spinors expressed in terms of $\sigma_\pm$ as the investigation of the KSAs in terms of the $\tau_\pm$ Killing spinors is similar.
Amongst the conditions (\ref{billI}), (\ref{billII}) and (\ref{billIII}) that we have put on the bilinears, the only relevant one is  (\ref{billII}).  This also follows
from the gravitino KSE as the connection has holonomy contained in the $Spin(8)$ group.  In particular one does not have to assume the compactness of the internal space or
use the maximum principle.

The KSAs  of the heterotic AdS$_3$ backgrounds can be easily constructed from first principles without using the super-Jacobi identities we have employed in the type II theories.  This is mainly due to the observation that the solution of the gravitino KSE puts strong restrictions
on Killing spinors namely that they should have a non-trivial isotropy group in $Spin(9,1)$. The anti-commutator of odd  generators can be read from the results of appendix C. Moreover,  the commutator of even and odd generators
can be easily found.  Indeed  upon using the gravitino KSE the spinorial Lie derivative of any Killing spinor $\epsilon$ with respect to any 1-form bilinear $X$ can be expressed as
\bea
{\cal L}_X\epsilon={1\over4} i_X \slashed {H} \epsilon~,
\label{hsld}
\eea
where $H$ is the 3-form field strength.  As both $\epsilon$ and $H$ are known in all cases, the right-hand-side of this equation can be easily evaluated.

\subsection{KSAs for heterotic backgrounds }

The KSAs of $N=2,\, 4$ and $N=6$  AdS$_3$ backgrounds can be either constructed from first principles as described in the previous section or can be read from the
results we have already presented  for the type II backgrounds.  In either case, they are unique and isomorphic to $\mathfrak{osp}(1\vert 2)$, $\mathfrak{osp}(2\vert 2)$
and $\mathfrak{osp}(3\vert 2)$, respectively.

 It remains to investigate the  $N=8$ AdS$_3$ backgrounds. As in type II theories, the Killing spinors are given  in (\ref{hks2}) but now $r=1,2,3, 4$.
 Furthermore, we express $K_{Ar, Bs}$  as in
(\ref{hbi2}) and there are potentially six  1-forms $\tilde K_{rs}=-\tilde K_{sr}$ along the internal space. The associated vector fields of these commute with
those of $K_{AB}$.  However, the Killing spinors of $N=8$  heterotic AdS$_3$ backgrounds are restricted to be  $\mathfrak{su}(2)$ invariant and such backgrounds admit only three  1-form bilinears along the internal space \cite{hetads}.  As a consequence only three of the six 1-forms $\tilde K_{rs}$ are linearly independent. This is imposed by
requiring that $\tilde K_{rs}$ is self-dual, ie
\bea
\tilde K_{rs}={1\over2} \epsilon_{rs}{}^{pq} \tilde K_{pq}~,
\eea
for some choice of ordering of Killing spinors.  The commutator $[\tilde V_{rs}, Q_{At}]$ can either be found  from explicitly computing
 the spinorial Lie derivative of $\tilde K_{rs}$   using (\ref{hsld}) and the form of the 3-form flux for such backgrounds given in \cite{hetads} or it can be read from
 the results for type II backgrounds as this case corresponds to the $\hat\alpha=-1$ case in (\ref{d21aa}). In either case
writing $\tilde V_{rs}= (\omega^{(+)}_P)_{rs} \tilde V_P$, $P=1,2,3$, where $\omega^{(+)}_P$ is a orthonormal  basis in the space of self-dual 2-forms in $\bR^4$ such that $[\omega^{(+)}_P, \omega^{(+)}_Q]=-2\epsilon_{PQ}{}^S \omega^{(+)}_S$, one finds that
\bea
&&[\tilde V_P, Q_{Ar}]= -\ell^{-1} (\omega^{(+)}_P)_r{}^s Q_{As}~,
\cr
&&[\tilde V_P, \tilde V_Q]=- \ell^{-1} \epsilon_{PQ}{}^S  \tilde V_S~.
\label{superh6}
\eea
This is a real form of the $\mathfrak{sl}(2\vert 2)/1_{4\times 4}$ superalgebra and the isometry algebra of the internal space is $\mathfrak{so}(3)$.  Central charges
do not arise in the heterotic case. The KSAs of AdS$_3$
heterotic backgrounds are tabulated in table 3.

\begin{table}[h]
 \caption{Heterotic AdS$_3$  KSAs }
 \begin{center}
 \begin{tabular}{|c|c|c|}
  \hline
  $N$ & $\mathfrak{g}$ & $\mathfrak{t}_0$
  \\ \hline
  2  &$\mathfrak{ osp}(1\vert 2)$ & -
  \\ \hline
  4  & $\mathfrak{osp}(2\vert 2)$ & $\mathfrak{so}(2)$
  \\ \hline
  6 & $\mathfrak{osp}(3\vert 2)$ & $\mathfrak{so}(3)$
  \\ \hline
  8 &$\mathfrak{ sl}(2\vert 2)/1_{4\times 4}$ & $\mathfrak{so}(3)$
  \\ \hline
 \end{tabular}
 \end{center}
\end{table}

\section{AdS$_4$ in D=11 and Type II Theories}

\subsection{$N=4$}

  AdS$_4$ backgrounds  preserve $4k$
supersymmetries so $N=4$ is the minimal case. Choosing the coordinates of  AdS$_4$ as  $(u,r,z,x)$,  the four Killing spinors (\ref{ksp1}) and (\ref{ksp2}) can be written as
\bea
 \epsilon_1 &=& \sigma_+~, \quad \epsilon_2 = A \Gamma_{- z} \sigma_+ + 2 \ell^{-1} u \sigma_+ - \ell^{-1} x A e^{z / \ell} \Gamma_{- x} \sigma_+
 \cr
 \epsilon_3 &= &e^{-z / \ell} \left( \Gamma_{z x} \sigma_+ - \ell^{-1} r A^{-1} \Gamma_{- x} \sigma_+ \right) - \ell^{-1} x \sigma_+~, \quad\epsilon_4 = A e^{z / \ell} \Gamma_{- x} \sigma_+~,
 \label{eq:minimal_ads4_killing_spinors}
\eea
where we have used   $\sigma_- = A \Gamma_{- z} \sigma_+$,  $\tau_+ = \Gamma_{z x} \sigma_+$ and  $\tau_- = A \Gamma_{- x} \sigma_+$.

The 1-form bilinears have been computed in appendix C.  They span all ten isometries of AdS$_4$.  Furthermore, all
1-form bilinears associated with isometries  of the internal space vanish.
The spinorial Lie derivatives of the Killing spinors along the isometries of AdS$_4$ can be easily extracted from the formulae in appendix C. It turns out that the resulting KSA  is
\bea
\{Q_A, Q_B\}=V_{AB}~,~~~[V_{AB}, Q_C]=-\ell^{-1} (\epsilon_{CA} Q_B+ \epsilon_{CB} Q_A)~,
\eea
where $A,B,C=1,\dots, 4$,   $\epsilon_{AB}$ is the symplectic invariant 2-form put into canonical form with $\epsilon_{12}=-\epsilon_{34}=1$ and the spinor $\sigma_+$ has been normalized as $2\parallel\sigma_+\parallel^2=1$.
This superalgebra is isomorphic to $\mathfrak{osp}(1\vert 4)$.

\subsection{$N=8$} \label{sec:m_iia_iib_ads4_extended}

The Killing spinors of $N=8$ AdS$_4$ backgrounds are generated by the choice of two linearly independent spinors $\sigma_+^r$, $r=1,2$.  As each $\sigma_+$ generates four Killing spinors as in (\ref{eq:minimal_ads4_killing_spinors}),
we shall denote the Killing spinors with $\epsilon_{Ar}$.  Without loss of generality  $\sigma_+^r$ can be chosen to be orthogonal
$2 \langle \sigma_+^r, \sigma_+^s\rangle =\delta_{rs}$. Furthermore  a direct application of the formulae in appendix C reveals that
\bea
\{Q_{Ar}, Q_{Bs}\}= \delta_{rs} V_{AB}+ \epsilon_{AB} \tilde V_{rs}~,
\label{ads4c1}
\eea
where  $\tilde V_{rs}=-\tilde V_{sr}$.  The generators $V_{AB}$ are associated to isometries along AdS$_4$ as in the minimal $N=4$ case and $\tilde V_{rs}$ is a generator associated to the 1-form
 bilinear
\bea
\tilde K_{rs} =2 A\langle \sigma^r_+, \Gamma_{zi} \sigma_+^s\rangle  \,\bbe^i~,
\label{ads41f}
\eea
which  gives rise to a Killing vector field along the internal space.
Note that $\tilde V_{rs}$ commutes with $ V_{AB}$.

A further direct computation using the formulae in appendix C also reveals that
\bea
[V_{AB}, Q_{Cr}]=-\ell^{-1} (\epsilon_{CA} Q_{Br}+ \epsilon_{CB} Q_{Ar})~.
\label{ads4c2}
\eea
It remains to determine the commutators $[\tilde V, Q]$.  For this, we shall use the closure of the KSA as an explicit computation will require further details
of the geometry of the internal space that are not available. As $\tilde K$ is along the internal directions it preserves the functional dependence of Killing spinors $\epsilon_{Ar}$  on AdS$_4$ coordinates.  So
 ${\cal L}_{\tilde K_{rs}} \epsilon_{At}= \beta_{rst}{}^\ell \epsilon_{A\ell}$ for some constants $\beta$.  Then one can compute  $[\tilde V, Q]$  using the  super-Jacobi identities.  In particular
  an argument similar to that employed for  the  AdS$_3$ $N>2$ backgrounds to prove that $\alpha$ is a 4-form, together with
(\ref{ads4c1}) and (\ref{ads4c2}) leads to
\bea
[\tilde V_{rs} , Q_{At}]=-\ell^{-1} (\delta_{tr} Q_{As}-\delta_{ts} Q_{Ar})~.
\label{ads4c3}
\eea
The KSA given by the (anti)-commutators (\ref{ads4c1}), (\ref{ads4c2}) and (\ref{ads4c3}) is isomorphic to $\mathfrak{osp}(2\vert 4)$.

\subsection{$N=12$}

The Killing spinors $\epsilon_{Ar}$ of these backgrounds are given as in  (\ref{eq:minimal_ads4_killing_spinors}) and depend on three spinors $\sigma_+^r$, $r=1,2,3$, which without loss
of generality we can  choose as $2 \langle \sigma_+^r, \sigma_+^s\rangle =\delta_{rs}$.

Using the results in appendix C, one can show that the anticommutators of the $Q_{Ar}$ generators are
\bea
\{Q_{Ar}, Q_{Bs}\}= \delta_{rs} V_{AB}+ \epsilon_{AB} \tilde V_{rs}~,
\label{ads4c112}
\eea
 where now the three generators
$\tilde V_{rs}$, $\tilde V_{rs}=-\tilde V_{sr}$, are associated with the three  1-form bilinears
\bea
\tilde K_{rs} =2 A \langle \sigma^r_+, \Gamma_{zi} \sigma_+^s\rangle  \,\bbe^i~,
\label{ads42f}
\eea
which in turn give rise to Killing vector fields in the internal space.  Similarly the commutator of $V$ and $Q$ generators is given as in (\ref{ads4c2}). It remains to determine
the commutators $[\tilde V_{rs}, Q_{At}]$.  Again we shall use the closure of KSA for this. If $t$ is equal to either $r$ or $s$, then a similar argument to that used for $N=8$ gives that the commutator is as
in (\ref{ads4c3}).  It remains to determine the commutators for $t\not= r,s$.  As $\tilde K_{rs}$ are along the internal space, the spinorial Lie derivative preserves the
functional dependence of the Killing spinors on the AdS$_4$ coordinates and so the commutator $[\tilde V_{rs}, Q_{At}]$ must close to  a linear combination
of $Q_{A\ell}$ generators. Because of this, it suffices to choose $A=1$ and consider first the case $r=1$, $s=2$ and $t=3$. Then applying the super-Jacobi identity using (\ref{ads4c112}),
one finds
\bea
[\tilde V_{12}, Q_{13}]=-[\{Q_{31}, Q_{42}\}, Q_{13}]=[\{Q_{13} , Q_{31}\}, Q_{42}]+[\{Q_{42}, Q_{13}\}, Q_{31}]=0~.
\label{66xx}
\eea
Thus $[\tilde V_{12}, Q_{A3}]=0$ as well.  A similar argument implies  that $[\tilde V_{rs}, Q_{At}]=0$ for $t\not=r,s$.  As a result the commutator is
\bea
[\tilde V_{rs} , Q_{At}]=-\ell^{-1} (\delta_{tr} Q_{As}-\delta_{ts} Q_{Ar})~.
\label{ads4c312}
\eea

The argument presented above to determine the commutator (\ref{ads4c312}) is general and does not depend of the range of the indices $r,s,t$. Clearly,
the KSAs of AdS$_4$ backgrounds with extended supersymmetry are more restricted than those AdS$_3$ backgrounds. The key reason for this is that unlike the
AdS$_3$ case,  the generators $\tilde V_{rs}$ can be written as anti-commutators of odd generators in two different ways for $A=1, B=2$ and for $A=3, B=4$.

The commutator $[\tilde V_{rs}, \tilde V_{r's'}]$ can be easily computed using the super-Jacobi identities to reveal that
\bea
[\tilde V_{rs}, \tilde V_{r's'}]&=&- \ell^{-1} \big( \delta_{rr'} \tilde V_{ss'}- \delta_{sr'}\tilde V_{rs'}-\delta_{rs'} \tilde V_{sr'}+\delta_{ss'} \tilde V_{rr'}\big)~,
\label{ads4c412}
\eea
ie the Lie algebra of the Killing vector fields of the internal space is $\mathfrak{so}(3)$. The KSA of AdS$_4$ backgrounds with $N=12$ supersymmetries is isomorphic to $\mathfrak{ osp}(3\vert 4)$.

\subsection{$N =16$}

The Killing spinors $\epsilon_{Ar}$ are again given as in (\ref{eq:minimal_ads4_killing_spinors}) but now determined by similarly normalized spinors $\sigma_+^r$ for $r=1,2,3,4$.
The squaring operation of the Killing spinors which gives  the anti-commutators $\{Q_{Ar}, Q_{Bs}\}$ leads to an expression as in (\ref{ads4c112}) but now for $r,s=1,2,3,4$.
The generators $\tilde V_{rs}$ are associated with Killing vectors along the internal space.
Also the commutator $[V_{AB}, Q_{Cr}]$ is given as in (\ref{ads4c2}).

The remaining commutators which need to be determined are  $[\tilde V_{rs}, Q_{At}]$.  If $t=r$ or $t=s$, an argument similar to the one produced for the $N=8$ case
leads to a commutator as in (\ref{ads4c3}).  On the other hand  if all $r,s,t$ are distinct, a similar argument to that used in the $N=12$ case implies that the
commutator vanishes. Thus the commutator is given as in (\ref{ads4c312}) but now for $r,s,t=1,2,3,4$.
 Furthermore as a consequence of this and the super-Jacobi identities the Lie algebra of $V_{rs}$ is as in (\ref{ads4c412}), i.e.~isomorphic to $\mathfrak{so}(4)$.
 Therefore the KSA of $N=16$ backgrounds is isomorphic to $\mathfrak{ osp}(4\vert 4)$.

\subsection{$N >16$}

It is straightforward to generalize the results we have obtained so far to all $N>16$ backgrounds. One can show that the KSA  of $N=4k$
backgrounds is $\mathfrak{ osp}(k\vert 4)$. The non-vanishing commutators are
\bea
\{Q_{Ar}, Q_{Bs}\}&=& \delta_{rs} V_{AB}+ \epsilon_{AB} \tilde V_{rs}~,
\cr
[V_{AB}, Q_{Cr}]&=&-\ell^{-1} (\epsilon_{CA} Q_{Br}+ \epsilon_{CB} Q_{Ar})~,
\cr
[\tilde V_{rs} , Q_{At}]&=&-\ell^{-1} (\delta_{tr} Q_{As}-\delta_{ts} Q_{Ar})~,
\cr
[\tilde V_{rs}, \tilde V_{r's'}]&=&- \ell^{-1} \big( \delta_{rr'} \tilde V_{ss'}- \delta_{sr'}\tilde V_{rs'}-\delta_{rs'} \tilde V_{sr'}+\delta_{ss'} \tilde V_{rr'}\big)~,
\eea
where $A,B,C=1,\dots,4$ and $r,s,t, r', s'=1,\dots, k$.
The algebra of the Killing vector fields of the internal space is  $\mathfrak{so}(k)$.

We remark that the KSAs of AdS$_4$ backgrounds are more restricted than those
of AdS$_3$ backgrounds. The critical argument that explains the reason for this is produced below (\ref{ads4c312}) and uses in an essential way the fact  that the generators of the isometries in
the internal space  $\tilde V_{rs}$ appear as  bilinears of both type $\sigma$ and type $\tau$ Killing spinors. Equivalently, the enhanced symmetry of the spacetime imposes more stringent conditions on the remaining (anti-) commutators  which are sufficient,
together with the super-Jacobi identities, to specify the KSAs including the commutators involving the generators of the isometries of the internal space.
This is the case for all KSAs, $\mathfrak{g}$, of  AdS$_n$ $n>3$ backgrounds provided that $\mathfrak{g}_0=\mathfrak{so}(n-1,2)\oplus \mathfrak{t}_0$.

\section{AdS$_5$ in D=11 and Type II Theories}

It has been shown in \cite{mads, iibads, iiaads} that warped AdS$_5$ backgrounds preserve $8k$ supersymmetries.
Unlike the AdS solutions  we have investigated so far, the  minimal $N=8$  AdS$_5$    backgrounds  exhibit
a non-trivial isometry along the internal space. Because of this the analysis is somewhat different.

\subsection{$N=8$}

To begin let us take  $(r,u,z, x,y)$ as the coordinates of AdS$_5$, where we have set $x^1=x$ and $x^2=y$. The eight Killing spinors of the minimal AdS$_5$ solution are generated by a single spinor $\sigma_+$ after applying the elementary operations
described in section \ref{elop}. It  follows that  if $\sigma_+$ is a Killing spinor, then $\Gamma_{xy} \sigma_+$ is also a Killing spinor.  As a result the Killing spinors can be organized
in two pairs of four spinors as $\epsilon_{AI}$ where $A=1,\dots, 4$ and $I=1,2$. The four $\epsilon_{A1}$ spinors are as those of the minimal AdS$_4$ backgrounds generated by $\sigma^1_+=\sigma_+$ and  $\tau^1_+=\Gamma_{zx} \sigma^1_+, \sigma^1_-=A\Gamma_{-z}\sigma^1_+,
\tau^1_-=A\Gamma_{-z}\tau^1_+$.  The remaining four Killing spinors $\epsilon_{A2}$ are generated by  $\sigma_+^2=\Gamma_{xy}\sigma_+$ after applying  the same elementary operations.   With these identifications, each of the $\epsilon_{A1}$ and $\epsilon_{A2}$ spinors generate a $\mathfrak {osp}(1\vert 4)$ superalgebra
as in the minimal AdS$_4$ case. Then  a direct substitution into  the vector bilinears of appendix C reveals that
\bea
\{Q_{AI}, Q_{BJ}\}=\delta_{IJ} V_{AB}+\epsilon_{IJ}{\mathring {W}}_{AB}+ \epsilon_{IJ}\epsilon_{AB} K~,
\label{ads5mee}
\eea
where $\mathring { W}_{AB} \epsilon^{AB}=0$, $\epsilon_{AB}$ is the $\mathfrak{sp}(4)$ invariant tensor defined as in the AdS$_4$ case with $\epsilon_{AB} \epsilon^{AC}=\delta_B^C$,
and $\epsilon_{IJ}=-\epsilon_{JI}$.
There are 16 linearly independent 1-form bilinears. Ten are associated to $V_{AB}$ and five are associated to ${\mathring {W}}_{AB}$. These span the
 fifteen  isometries of AdS$_5$. $K$ on the other hand is associated with  an isometry along the internal space generated by 1-form bilinear
$K=2A \langle\sigma_+, \Gamma_{xyzi}\sigma_+\rangle \,\bbe^i$. The Lie algebra of the isometries is $\mathfrak{so}(2,4)\oplus \mathfrak{so}(2)$.

It remains to compute the commutators of $V$, $\mathring {W}$ and $K$ with the $Q$'s.  The commutators of $V$ and  $\mathring {W}$ with the $Q$'s can be read
from the spinorial derivatives on the Killing spinors presented in appendix C as these even generators are associated with isometries of AdS$_5$.  In particular
it is straightforward to see that
\bea
[V_{AB}, Q_{CI}]&=&-\ell^{-1} (\epsilon_{CA} Q_{BI}+ \epsilon_{CB} Q_{AI})~.
\label{ads5meo1}
\eea
It remains
to find the commutator of $K$ with the $Q$'s. For this observe that
\bea
\{Q_{11}, Q_{22}\}=-\{Q_{12}, Q_{21}\}=M+K~,~~~\{Q_{31}, Q_{42}\}=-\{Q_{32}, Q_{41}\}=M-K~,
\eea
where the generator $M$ is associated with the 1-form bilinear $-\ell^{-1} M^{xy}$.  We have used the normalization $2\parallel\sigma_+\parallel^2=1$.
Thus for every $Q_{AI}$ there is another odd generator $Q_{A'I'}$ such that $\{Q_{AI}, Q_{A'I'}\}\propto M\pm K$. This gives
 \bea
 [M\pm K, Q_{AI}]\propto [\{Q_{AI}, Q_{A'I'}\}, Q_{AI}]=-{1\over2} [\{Q_{AI}, Q_{AI}\},  Q_{A'I'}]=\ell^{-1} \epsilon_{A'A} Q_{AI'}~.
 \eea
 As the commutators $[M, Q_{AI}]$ are known, one can find all the commutators $[K, Q_{AI}]$.  These results can be collected as
\bea
[W_{AB}, Q_{CI}]&=&-\ell^{-1} \epsilon_{IJ}(\epsilon_{CA} Q_{BJ}- \epsilon_{CB} Q_{AJ}+2 \epsilon_{AB} Q_{CJ})~,
\label{ads5meo2}
\eea
where $W_{AB}=\mathring { W}_{AB}+\epsilon_{AB} K$. In particular, one has
\bea
[K, Q_{AI}]=-{3\over2} \ell^{-1} \epsilon_{IJ} Q_{AJ}~.
\eea
Clearly the generator $K$ cannot be set to zero establishing that the internal spaces of all such backgrounds must have a non-trivial isometry.  The (anti)-commutators (\ref{ads5mee}), (\ref{ads5meo1}) and (\ref{ads5meo2}) determine the
KSA and it is isomorphic to $\mathfrak{sl}(1\vert 4)$.

\subsection{$N=16$}

As the AdS$_5$ backgrounds preserve $8 k$ supersymmetries, the next case to investigate is $N=16$.  For this set $\sigma_+^1=\sigma_+$, where $\sigma_+$ is the
Killing spinor of the $N=8$ case and introduce another Killing spinor $\sigma_+^2$ which is linearly independent from both $\sigma_+^1$ and $\Gamma_{xy}\sigma^1_+$.  Without loss of generality one can choose $\sigma_+^2$ to be orthogonal to  both $\sigma_+^1$ and $\Gamma_{xy}\sigma^1_+$.  As a result $\sigma_+^1$, $\Gamma_{xy}\sigma^1_+$, $\sigma_+^2$ and $\Gamma_{xy}\sigma^2_+$
can be chosen as mutually orthogonal.

 A direct inspection of the bilinears in appendix C reveals that the $Q$ anti-commutators can be arranged as
  \bea
\{Q^r_{AI}, Q^s_{BJ}\}=\delta_{IJ} \delta^{rs} V_{AB}+\epsilon_{IJ} \delta^{rs} \mathring{W}_{AB}+\delta_{IJ} \epsilon_{AB} \tilde V^{rs}+ \epsilon_{IJ} \epsilon_{AB} K^{rs}
\label{ads5qq16}
\eea
where the $V$ and $ \mathring{W}$ generators are as in the $N=8$ case while the generators $\tilde V^{rs}=-\tilde V^{sr}$ and  $K^{rs}=K^{sr}$ are associated to the
1-form bilinears
\bea
 &&K^{rs}=2A \langle\sigma^r_+, \Gamma_{xyzi}\sigma^s_+\rangle \,\bbe^i~,~~~ \tilde V^{rs}=2A \mathrm{Re}\, \langle\sigma^r_+, \Gamma_{zi}\sigma^s_+\rangle \,\bbe^i~,
 \label{ads5inbi}
  \eea
and $r,s=1,2$.

It remains to investigate the commutators of even and odd generators. All such commutators that involve generators of isometries   along the AdS$_5$ can be easily computed using the spinorial Lie derivatives. In particular, one has
\bea
[\mathring {W}_{AB}, Q^r_{CI}]&=&-\ell^{-1} \epsilon_{IJ}(\epsilon_{CA} Q^r_{BJ}- \epsilon_{CB} Q^r_{AJ}+{1\over2} \epsilon_{AB} Q^r_{CJ})~.
\label{ads5wq16}
\eea
It remains to compute the commutators of  generators of isometries  along the internal directions with the odd generators.
To find the commutator  $ [\tilde V^{rs}, Q^t_{AI}]$,   observe that for $I=J$ the anti-commutator (\ref{ads5qq16}) reduces to that of $N=8$ AdS$_4$ backgrounds.
Then a similar argument to that produced in the AdS$_4$ case leads to the commutator
\bea
[\tilde V^{rs}, Q_{AI}^t]&=&-\ell^{-1} (\delta^{tr}  Q_{AI}^s-\delta^{ts}  Q_{AI}^r)~.
\label{ads5vq16}
\eea
To find the commutators of  $[K^{rs},Q^t_{AI}]$, observe that
  \bea
  K^{rs}+\mathring{W}_{12} \delta^{rs}=\{Q^r_{11}, Q^s_{22}\}=-\{Q^r_{12}, Q^s_{21}\}~,
  \cr
 K^{rs}-\mathring{W}_{34} \delta^{rs}=-\{Q^r_{31}, Q^s_{42}\}=\{Q^r_{32}, Q^s_{41}\}~.
  \eea
  Therefore for each $Q^t_{AI}$, there are $Q^r_{A'1}$ and $Q^s_{B'2}$ such that $\epsilon_{AA'}=\epsilon_{AB'}=0$ and  $ K^{rs}$ appears in the anti-commutator $\{Q^r_{A'1}, Q^s_{B'2}\}$.
  Using this we have
  \bea
  [K^{rs}\epsilon_{A'B'}+ \delta^{rs} \mathring{W}_{A'B'}, Q_{AI}^t] &\propto& [\{Q^r_{A'1}, Q^s_{B'2}\}, Q_{AI}^t]=- [\{ Q_{AI}^t, Q^r_{A'1}\},  Q^s_{B'2}]
  \cr &&-
  [\{Q^s_{B'2},  Q_{AI}^t\}, Q^r_{A'1} ]~.
  \eea
 Observe that the right-hand-side of the equation above does not depend on  generators $K^{rs}$  and  all the  (anti-)commutators are known. Thus  one can use the above
  formula to find
\bea
[K^{rs}, Q^t_{AI}]=-\ell^{-1} \epsilon_{IJ} (\delta^{tr} Q^s_{AJ}+\delta^{ts} Q^r_{AJ}-{1\over2} \delta^{rs} Q^t_{AJ})~.
\label{ads5kq16}
\eea
This superalgebra defined by the (anti-)commutators (\ref{ads5qq16}), (\ref{ads5wq16}), (\ref{ads5vq16}) and (\ref{ads5kq16}) is isomorphic to  $\mathfrak{sl}(2\vert 4)$ and the isometry algebra of the transverse space is $\mathfrak{u}(2)$.

\subsection{Extended supersymmetry}

The two remaining cases to investigate are for  $N=24$ and for $N=32$.  The bilinears that lie along the internal directions are again given as in (\ref{ads5inbi}) but now
for $r,s=1,2,3$ and $r,s=1,\dots,4$, respectively.  The anti-commutator of the $Q$ generators is again given in (\ref{ads5qq16}) but now either  $r,s=1,2,3$ or $r,s=1,\dots,4$, where again
the generators $\tilde V^{rs}$ and $K^{rs}$ are associated to bilinears that lie along the internal space.
The  commutators $[\tilde V^{rs}, Q^t_{AI}]$ and  $[K^{rs}, Q^t_{AI}]$ are again given by  (\ref{ads5vq16}) and  (\ref{ads5kq16}), respectively, for either  $r,s=1,2,3$ or $r,s=1,\dots,4$.  This
is because the argument to establish (\ref{ads5vq16}) and  (\ref{ads5kq16}) for  the $N=16$ case is  not sensitive to the range of the indices $r,s$ and $t$.
 The resulting superalgebra in the $N=24$ case is isomorphic to  $\mathfrak{sl}(3\vert 4)$ and the Lie algebra of  isometries of the internal space is  $\mathfrak{u}(3)$. However, although this local analysis allows for the existence of $N=24$ solutions, a global analysis given in \cite{sbjggp} which makes use of a maximum principle argument
on the (compact and without boundary) internal space, excludes such solutions.

  In the $N=32$ case observe
 that the generator $C={1\over 4} \delta_{rs} K^{rs}$ is central as it commutes with all the $Q$'s and all the even generators. So either $C$ does not appear in the right-hand-side
 of the anti-commutator (\ref{ads5qq16}) in which case the superalgebra is isomorphic to $\mathfrak{sl}(4\vert 4)/\lambda 1_{8\times 8}$ and  the isometry algebra of the internal space is  $\mathfrak{su}(4)=
 \mathfrak{so}(6)$ or the KSA is not simple. The latter case does not occur as the only maximally supersymmetric AdS$_5$ background is the IIB AdS$_5\times S^5$  solution and the effective isometry
 algebra of the transverse space  is $\mathfrak{so}(6)$.  However this cannot  be  deduced    on  symmetry considerations alone as the classification of maximal
 supersymmetric
 solutions of IIB theory is also required \cite{maxsusy}.

\section{AdS$_6$ in D=11 and Type II Theories}

AdS$_6$ backgrounds preserve either 16 or 32 supersymmetries.  It has been known for some time that there are no AdS$_6$ backgrounds with 32 supersymmetries in 11-dimensional
and type II supergravities  \cite{maxsusy}.  As a result the only case that remains to be investigated is that of 16 supersymmetries.  For this, let us denote the coordinates of AdS$_6$ with $(u,r,z,x^a)$, $a=1,2,3$, and introduce the odd generators
\bea
Q_A=Q_{\epsilon_A(\sigma_+)}~,~~Q_{Aa}=Q_{{1\over2} \epsilon_a{}^{bc}
\epsilon_A(\Gamma_{bc}\sigma_+)}~,~~\tilde Q_A=Q_{\epsilon_{A+2}(\Gamma_{z123}\sigma_+)}~,~~\tilde Q_{Aa}=Q_{\epsilon_{A+2} (\Gamma_{za}\sigma_+)}
\eea
where  we have used the expression  of Killing spinors in (\ref{ksp1}) together with $\sigma_-=A \Gamma_{-z} \sigma_+$
and $\tau_-=A \Gamma_{-z} \tau_+$ as well as the relation between $\tau_+$ and $\sigma_+$ spinors. The $a, b, c$ indices are raised with respect to the flat metric and $\epsilon_{abc}$
is the Levi-Civita tensor.

A direct substitution of the Killing spinors into
the 1-form bilinears of appendix C reveals that the non-vanishing anti-commutators are
\bea
&&\{Q_A, Q_B\}= V_{AB}~,~~\{Q_A, Q_{Ba}\}={1\over2}\epsilon_{AB} \epsilon_a{}^{bc} K^{(-)}_{bc}~,~~\{Q_{Aa}, Q_{Bb}\}=\delta_{ab} V_{AB}+\epsilon_{AB} K^{(+)}_{ab}
\cr
&&\{\tilde Q_A, \tilde Q_B\}= \tilde V_{AB}~,~~\{\tilde Q_A, \tilde Q_{Ba}\}={1\over2}\epsilon_{AB} \epsilon_a{}^{bc} K^{(-)}_{bc}~,~~\{\tilde Q_{Aa}, \tilde Q_{Bb}\}=\delta_{ab} \tilde V_{AB} -\epsilon_{AB} K^{(+)}_{ab}
\cr
&&\{Q_A, \tilde Q_{Ba}\}=\{\tilde Q_B, Q_{Aa}\}=V_{AB,a}~,~~~\{Q_{Aa}, \tilde Q_{Bb}\}=-\epsilon_{ab}{}^c V_{AB,c}
\eea
The generators $V_{AB}$ and $\tilde V_{AB}$ are associated to 1-form bilinears (\ref{kvf2}) and (\ref{tauiso}), respectively. The generators $K^{(\pm)}_{ab}$ are associated
to 1-form bilinears
\bea
\pm \ell^{-1} M_{ab}+ 2A \langle \sigma_+, \Gamma_{zi} \Gamma_{ab} \sigma_+\rangle\, \bbe^i
\eea
and the generators $V_{AB,a}$ are associated to the bilinears
\bea
V_{11,a}=-\ell^{-1} M^{-}{}_{a}~,~~V_{12,a}=-\lambda^a~,~~V_{21,a}=2\ell^{-1}M^{z}{}_{a}+\lambda_a~,~~V_{22,a}=2\ell^{-1} M^{+}{}_{a}
\eea
where $2 \parallel\sigma_+\parallel^2=1$. There are at most 3 Killing vectors along the internal space associated with
the bilinears
\bea
K_{ab}=2 A\langle \sigma_+,  \Gamma_{zi} \Gamma_{ab} \sigma_+\rangle \,\bbe^i~,~~~~a,b=1,2,3~.
\eea
Note that the bilinear $K_{abc}=2A\langle \sigma_+, \Gamma_{abc} \Gamma_i \sigma_+\rangle \,\bbe^i$ vanishes as a consequence of the conditions (\ref{billI}) in section \ref{consection}.

As in previous cases, to determine  commutators of the KSA it suffices to find the commutators of the generators $K_{ab}$ of the isometries of the internal space  with
$Q$'s  as the rest follow the explicit formulae in appendix C via the evaluation of spinorial Lie derivatives of the Killing spinors along the isometries of AdS$_6$.
To find $[K_{bc}, Q_{Aa}]$ and $[K_{bc}, Q_{A}]$ observe that
\bea
K^{(+)}_{ab}=-{1\over2} \epsilon^{AB} \{\tilde Q_{Aa}, \tilde Q_{Bb}\}~,
\eea
and so
\bea
[K_{bc}, Q_{Aa}]=-\ell^{-1} [M_{bc}, Q_{Aa}]+{1\over2} \epsilon^{BC}\big( [\{Q_{Aa}, \tilde Q_{Bb}\}, \tilde Q_{Cc}]+[\{\tilde Q_{Cc} , Q_{Aa} \}, \tilde Q_{Bb}]\big)~.
\eea
All even generators in the right-hand-side are associated with the isometries of AdS$_6$. As a result, the right-hand-side can be
found using the spinorial Lie derivatives of appendix C.  This determines the commutator $[K_{bc}, Q_{Aa}]$.  A similar argument also determines
all the other commutators of $K_{ab}$ with the remaining $Q$'s and $\tilde Q$'s.  In particular all the non-vanishing commutators of even with odd generators are
\bea
&&[V_{AB}, Q_{C\alpha}]=-\ell^{-1} (\epsilon_{CA} Q_{B\alpha}+\epsilon_{CB} Q_{A\alpha})~,~~[\tilde V_{AB}, \tilde Q_{C\alpha}]=\ell^{-1} (\epsilon_{CA} \tilde Q_{B\alpha}+\epsilon_{CB} \tilde Q_{A\alpha})~,~~
\cr
&&[V_{AB, a}, Q_C]= \ell^{-1} \epsilon_{AC} \tilde Q_{Ba}~,~~~[V_{AB, a}, Q_{Cb}]= \ell^{-1} \epsilon_{AC} (\delta_{ab}\tilde Q_B-\epsilon_{ab}{}^c \tilde Q_{Bc})~,
\cr
&&[V_{AB, a}, \tilde Q_C]= -\ell^{-1} \epsilon_{BC} Q_{Aa}~,~~~[V_{AB, a}, \tilde Q_{Cb}]= -\ell^{-1} \epsilon_{BC} (\delta_{ab} Q_A+\epsilon_{ab}{}^c  Q_{Ac})~,
\cr
&&
[K_{ab},  Q_A]=-{3\over2}\ell^{-1} \epsilon_{ab}{}^c  Q_{Ac}~,~~~[K_{ab},  Q_{Ac}]={3\over2} \ell^{-1} \epsilon_{abc}  Q_A+{3\over2} \ell^{-1}
( Q_{Aa} \delta_{bc}-  Q_{Ab} \delta_{ac})~,
\cr
&&[M_{ab},  Q_A]=-{1\over2} \epsilon_{ab}{}^c  Q_{Ac}~,~~~[M_{ab}, \ Q_{Ac}]={1\over2} \epsilon_{abc} Q_A-{1\over2}( Q_{Aa} \delta_{bc}- Q_{Ab} \delta_{ac})~,
\cr
&&
[K_{ab}, \tilde Q_A]={3\over2}\ell^{-1} \epsilon_{ab}{}^c \tilde Q_{Ac}~,~~~[K_{ab}, \tilde Q_{Ac}]=-{3\over2} \ell^{-1} \epsilon_{abc} \tilde Q_A+{3\over2} \ell^{-1}
(\tilde Q_{Aa} \delta_{bc}- \tilde Q_{Ab} \delta_{ac})~,
\cr
&&[M_{ab}, \tilde Q_A]={1\over2} \epsilon_{ab}{}^c \tilde Q_{Ac}~,~~~[M_{ab}, \tilde Q_{Ac}]=-{1\over2} \epsilon_{abc} \tilde Q_A-{1\over2}(\tilde Q_{Aa} \delta_{bc}- \tilde Q_{Ab} \delta_{ac})~,
\eea
where $Q_{A\alpha}=(Q_A, Q_{Aa})$ and similarly for $\tilde Q_{A\alpha}$.
The KSA is isomorphic to $\mathfrak{f}^*(4)$ a real form of  $\mathfrak{f}(4)$ with $\mathfrak{f}^*(4)_0=\mathfrak{so}(5,2)\oplus \mathfrak{so}(3)$.
The Lie subalgebra generated by $K_{ab}$ is isomorphic to $\mathfrak{so}(3)$. The generators $K_{ab}$ cannot be set to zero as this will violate the super-Jacobi identities. So one expects that all these backgrounds admit an effective $\mathfrak{so}(3)$ action.

\section{AdS$_7$ in D=11 and Type II Theories}

To begin, a minimally supersymmetric AdS$_7$ background admits sixteen Killing spinors and   the odd generators can be identified  as
\bea
Q_A=Q_{\epsilon_A(\sigma_+)}~,~~Q_{Ar}={1\over2}\omega^{(-) ab}_rQ_{\epsilon_A(\Gamma_{ab}\sigma_+)}~,~~\tilde Q_{Aa}=Q_{\epsilon_{A+2}(\Gamma_{za}\sigma_+)}~,~~
\eea
where $\omega^{(-)}_r$ is a basis of anti-self-dual 2-forms in $\bR^4$ with
\bea
\omega^{(-)}_{r ab} \omega_s^{(-)b}{}_{ c} =-\delta_{rs} \delta_{ac}-\epsilon_{rs}{}^t \omega^{(-)}_{t ac}~.
\eea
Observe that there are only 3 independent generators $Q_{Aab}$ for each $A$ as $\sigma_+$ is restricted to satisfy\footnote{One can also choose
$\Gamma_{abcd}\sigma_+=-\epsilon_{abcd} \sigma_+$ and this case can be treated in a similar way.} $\Gamma_{abcd}\sigma_+=\epsilon_{abcd} \sigma_+$.
Computing the 1-form bilinears, we find that
\bea
&&\{Q_A, Q_{A'}\}=V_{AA'}~,~~~\{Q_A, Q_{Br}\}=\epsilon_{AB} (-2\ell^{-1} M_r+K_r) ~,~~~
\cr
&&\{Q_{Ar}, Q_{A's}\}=4 V_{AA'} \delta_{rs}+2 \epsilon_{AA'} \epsilon_{rs}{}^t (-2\ell^{-1} M_t+K_t)~,
\cr
&&\{\tilde Q_{Aa}, \tilde Q_{A'a'}\}=\tilde V_{AA'} \delta_{aa'}-\epsilon_{AA'}( \ell^{-1} M^{(+)}_{aa'}+ K_{aa'})~,
\cr
&&\{Q_A, \tilde Q_{Bb}\}= V_{AB,b}~,~~~\{Q_{Ar}, \tilde Q_{Bb}\}=2 V_{AB, c}\, \omega^{(-)}_r{}^c{}_b~,
\eea
where
\bea
K_{aa'}=2 A \langle \sigma_+, \Gamma_{zi} \Gamma_{aa'}\sigma_+\rangle~,~~~M_r={1\over2} \omega^{(-)}_r{}^{ab} M_{ab}~,~~~M^{(\pm)}_{ab}= M_{ab}\pm{1\over2} \epsilon_{ab}{}^{cd} M_{cd}~,
\eea
and
\bea
V_{11,a}=-\ell^{-1} M^-{}_a~,~~~V_{12,a}=-\lambda_a~,~~~V_{21,a}=2\ell^{-1} M^z{}_a+\lambda_a~,~~~V_{22,a}=2\ell^{-1} M^+{}_a~.
\eea

\begin{table}[h]
\begin{center}
\vskip 0.3cm
\underline {  AdS$_k$ KSAs in $d=10$ and $d=11$}
 \vskip 0.3cm
 \begin{tabular}{|c|c|c|c|c|c|}
  \hline
  $N$ & AdS$_4$ & AdS$_5$ & AdS$_6$ & AdS$_7$
  \\ \hline
  4  &$\mathfrak{ osp}(1\vert 4)$ & -       & -    & -
  \\ \hline
  8  & $\mathfrak{osp}(2\vert 4)$ & $\mathfrak{sl}(1\vert 4) $& -    & -
  \\ \hline
  12 &$\mathfrak{ osp}(3\vert 4)$ & -       & -    & -
  \\ \hline
  16 &$ \mathfrak{osp}(4\vert 4)$ & $\mathfrak{ sl}(2\vert 4)$ &$\mathfrak{ f}^*(4) $& $\mathfrak{osp}(6,2\vert 2)$
  \\ \hline
  20 &$\mathfrak{ osp}(5\vert 4)$ & -       & -    & -
  \\ \hline
  24 &$\mathfrak{ osp}(6\vert 4)$ &$ \mathfrak{sl}(3\vert 4)$ & -    & -
  \\ \hline
  28 &$\mathfrak{ osp}(7\vert 4)$ & -       & -    & -
  \\ \hline
  32 & $\mathfrak{osp}(8\vert 4)$ & $\mathfrak{ sl}(4\vert 4)/1_{8\times 8}$ & -    & $\mathfrak{osp}(6,2\vert 4)$
  \\ \hline
 \end{tabular}
 \end{center}
 \caption{In all cases, for AdS$_k$ backgrounds $\mathfrak{g}_0=\mathfrak{so}(k-1,2)\oplus \mathfrak{t}_0$. $\mathfrak{ f}^*(4) $ is a different
 real form to $\mathfrak{ f}(4) $  which appears in the AdS$_3$ case. }
\end{table}

The commutators of the even with the odd generators are
\bea
&&[V_{AB}, Q_{Ca}]=-\ell^{-1} (\epsilon_{CA} Q_{B\alpha}+\epsilon_{CB} Q_{Aa})~,~~[\tilde V_{AB}, \tilde Q_{Ca}]=\ell^{-1} (\epsilon_{CA} \tilde Q_{Ba}+\epsilon_{CB} \tilde Q_{Aa})~,~~
\cr
&&[V_{AA', a}, Q_B]=\ell^{-1} \epsilon_{AB} \tilde Q_{A'a}~,~~~[V_{AA', a}, Q_{Br}]=2\ell^{-1} \epsilon_{AB} \omega_r^{(-)}{}_a{}^b \tilde Q_{A'b}~,
\cr
&&[V_{AA', a}, \tilde Q_{Bb}]=-\ell^{-1} \epsilon_{ A'B} (\delta_{ab} Q_A+{1\over2} \omega^{(-)r}_{ab} Q_{Ar})~,~~~[K_r, Q_C]=-2\ell^{-1} Q_{Cr}~,
\cr
&&[M_{ab}, Q_A]=-{1\over4} \omega^{(-)}_r{}_{ab} Q_{Ar}~,~~~[M_{ab}, Q_{Ar}]=\omega^{(-)}_r{}_{ab} Q_A+{1\over2} \epsilon_r{}^{st} \omega^{(-)}_s{}_{ab} Q_{At} ~,~~~
\cr
&& [M_{ab}, \tilde Q_{Ac}]={1\over2} (\delta_{ca} \tilde Q_{Ab}-\delta_{cb} \tilde Q_{Aa})+{1\over2} \epsilon_{abc}{}^d \tilde Q_{Ad}~,
\cr
&&[K_r, Q_{As}]=8\ell^{-1} Q_A \delta_{rs}-4\ell^{-1} \epsilon_{rs}{}^t Q_{Ct}~,~~~[K_r, \tilde Q_{Aa}]=-4\ell^{-1} \omega^{(-)}_r{}_a{}^b \tilde Q_{Ab}~.
\eea
The commutators of $K_r$ with the odd generators have been found using a similar argument to that of the AdS$_6$ backgrounds.  The Lie algebra
of the $K_r$ generators is $\mathfrak{so}(3)$. The KSA is isomorphic to $\mathfrak{osp}(6,2 \vert 2)$.

It remains to investigate the maximally supersymmetric AdS$_7$ backgrounds.  It has been shown that all such solutions are locally isometric
to the AdS$_7\times S^4$ background in \cite{maxsusy}.  It can be shown using the technique illustrated above that the KSA  is isomorphic to
$\mathfrak{osp}(6,2\vert 4)$.  The calculation is very similar with the only difference that $\sigma_+$ is not restricted to be (anti-) chiral
with respect to the $\Gamma_{1234}$ chirality operator.

\begin{table}[h]

\begin{center}
\vskip 0.3cm
\underline { Isometry algebras of internal space}
 \vskip 0.3cm
 \begin{tabular}{|c|c|c|c|c|c|}
  \hline
  $N$ & AdS$_4$ & AdS$_5$ & AdS$_6$ & AdS$_7$
  \\ \hline
  4  & 0 & - & - & -
  \\ \hline
  8  & $\mathfrak{so}(2)$ & $\mathfrak{u}(1)$ & - & -
  \\ \hline
  12 & $\mathfrak{so}(3)$ & - & - & -
  \\ \hline
  16 & $\mathfrak{so}(4)$ & $\mathfrak{u}(2)$ & $\mathfrak{so}(3)$ & $\mathfrak{so}(3)$
  \\ \hline
  20 & $\mathfrak{so}(5)$ & - & - & -
  \\ \hline
  24 & $\mathfrak{so}(6)$ & $\mathfrak{u}(3) $ & - & -
  \\ \hline
  28 & $\mathfrak{so}(7)$ & - & - & -
  \\ \hline
  32 & $\mathfrak{so}(8)$ & $\mathfrak{su}(4)$ & - & $\mathfrak{so}(5)$
  \\ \hline
 \end{tabular}
 \end{center}
 \caption{For the maximally supersymmetric AdS$_5$ solution the isometry algebra of the internal space is $\mathfrak{su}(4)$ instead of $\mathfrak{u}(4)$ as the
 $\mathfrak{u}(1)$ generator does not act effectively on the transverse 5-sphere. }
\end{table}

\newpage

\section{Conclusions}

We have identified the KSAs, $\mathfrak{g}$, of all warped $AdS_k\times_w M^{d-k}$, $k\geq 3$, backgrounds with the most general allowed fluxes in $d=10$ and $d=11$ dimensions,  for which the even subalgebra $\mathfrak{g}_0$ decomposes into a direct sum of the isometries of AdS$_k$ and those
of the internal space $M^{d-k}$, $\mathfrak{g}_0=\mathfrak{so}(k-1,2)\oplus \mathfrak{t}_0$. The proof utilizes  (i) the solution of the KSEs for AdS backgrounds presented in \cite{mads, iibads, iiaads, hetads} and (ii) the closure
of the KSAs demonstrated in \cite{11jose, iibjose}.  Our   results are tabulated in tables 2, 3, 4 and 5.

We have demonstrated that the classification of AdS$_3$ KSAs is closely related to the classification of groups acting effectively and transitively
on spheres.  This is because the Lie algebra of isometries of the internal space, $\mathfrak{t}_0$,  is associated to a group that acts transitively on a sphere in the odd subspace $\mathfrak{g}_1$ of the KSA.  The classification of such groups  is a classic problem in geometry that has been solved some time ago \cite{ms} and it has been applied in \cite{simons} to simplify
the Berger classification of the holonomy groups of simply connected irreducible Riemannian manifolds. The KSAs of AdS$_3$ backgrounds may not be simple as they can exhibit
central generators. There are several potential KSAs for AdS$_3$ backgrounds for a given number of supersymmetries $N$. For the rest of AdS$_k$, $k>3$, backgrounds, we find that the KSAs are all classical  and they can be uniquely characterized by the pair $(k, N)$, i.e.~the AdS$_k$
space under investigation and the number of supersymmetries preserved by the background.

In the context of AdS$_k$/CFT$_{k-1}$, for $k>3$,  we have shown that the KSAs of all  AdS$_k$ backgrounds  which decompose as $\mathfrak{g}_0=\mathfrak{so}(k-1,2)\oplus \mathfrak{t}_0$
coincide with the expected superconformal algebras of  field theories.  So potentially all such backgrounds can  have a CFT$_{k-1}$ dual.
 The only exception perhaps is the KSA of the maximally supersymmetric AdS$_5$ background that can exhibit a central term which however
 vanishes in supergravity.
 In AdS$_3$ backgrounds,  there are many  KSAs  that  arise for a given number of supersymmetries $N$ and  can exhibit one or more central
generators.  The role of these central terms should be clarified  in both supergravity and in quantum theory.

Our results  have applications in the classification of supersymmetric AdS backgrounds.
If AdS backgrounds  preserve more than 16 supersymmetries, $N> 16$,  then under   some mild assumptions it can be shown that
the warp factor is constant and therefore they are products $AdS_k\times M^{d-k}$.  Furthermore they must be homogenous  \cite{hustler1} and so  $M^{d-k}=G/K$. The identification of all  KSAs of AdS$_k$ backgrounds allows to set  $\mathfrak{Lie}\,G=\mathfrak{t}_0$.
As all the Lie algrebras  $\mathfrak{t}_0$ of isometries of the internal spaces are known, all the internal spaces can be identified as homogenous spaces of groups with Lie
algebra $\mathfrak{t}_0$. So far, the AdS$_k$ backgrounds that preserve more than 16 supersymmetries have been classified for $k=4$ \cite{ahslgp} and  $k=5$ \cite{sbjggp}. In the former
case the classification of the KSAs for AdS$_4$ backgrounds has been utilized in an essential way.
 For $k>5$, AdS$_k$ backgrounds preserve   either 16 or 32 supersymmetries and so for $N>16$ are included in the classification of  maximal supersymmetric backgrounds  in \cite{maxsusy}.

For AdS$_k$ backgrounds that preserve 16 or less supersymmetries, $N\leq16$,  the KSA
 may not act transitively on the internal space.  Further progress on the classification of such backgrounds will require  a  detailed analysis of the orbits of the KSAs in the internal spaces
 as presented in \cite{jggpads6} for AdS$_6$ backgrounds.  Investigations of the geometry of such  AdS backgrounds  based of superalgebra considerations have been made  before, see e.g.~\cite{lin, hoker}. However now this can be done more systematically as all possibilities have been identified.


\vskip 1cm

\noindent{\bf Acknowledgements} \vskip 0.1cm
GP wishes to thank Jose Figueroa-O'Farrill and Alessandro Tomasiello for many helpful discussions.  UG and GP would like to thank MITP for providing a stimulating environment during the
workshop ``Geometry,  Gravity and Supersymmetry'' where part of this project was completed. GP is partially supported from the  STFC rolling grant ST/J002798/1. JG is supported by the STFC Consolidated Grant ST/L000490/1. UG is supported by the Swedish Research Council.

\vskip 0.5cm
\noindent{\bf Data Management:} \vskip 0.1cm
\noindent  No data beyond those presented and cited in this work are needed to validate this study.
\vskip 0.5cm

\setcounter{section}{0}
\setcounter{subsection}{0}
\setcounter{equation}{0}

\appendix{Invariance  of (massive) IIA fluxes }
\label{preservation}

In this appendix we will give a proof to the statement  that the {\sl Killing vector bilinears leave invariant all the fields of (massive) IIA supergravity}, i.e.~they are Killing vectors
 and preserve  all the fluxes. The proof will rely on the Killing spinor equations
\be\label{gkse}
{\cal D}_M \ep &\equiv& \nabla_M \ep + \tfrac{1}{8} H_{MP_1 P_2}\Gamma^{P_1 P_2}\Gamma_{11}\ep +\tfrac{1}{8} e^\Phi \tilde S\Gamma_M \ep \notag \\
&& +\tfrac{1}{16}e^\Phi \tilde F_{P_1 P_2}\Gamma^{P_1 P_2}\Gamma_M \Gamma_{11} \ep +\tfrac{1}{8\cdot 4!}e^\Phi \tilde G_{P_1 \cdots P_4}\Gamma^{P_1 \cdots P_4}\Gamma_M \ep
\ee
and
\be\label{akse}
{\cal A} \epsilon &\equiv & \partial_P \Phi \Gamma^P \ep + \tfrac{1}{12} H_{P_1 P_2 P_3}\Gamma^{P_1 P_2 P_3}\Gamma_{11}\ep +\tfrac{5}{4} e^\Phi \tilde S \ep \notag \\
&& +\tfrac{3}{8}e^\Phi\tilde F_{P_1 P_2}\Gamma^{P_1 P_2}\Gamma_{11}\ep +\tfrac{1}{4\cdot 4!}e^\Phi \tilde G_{P_1 \cdots P_4}\Gamma^{P_1 \cdots P_4} \ep ~,
\ee
where $\nabla$ is the spin connection, $H$ is the NS-NS 3-form field strength, $\tilde S, \tilde F, \tilde G$ are the RR $k$-form field strengths, for $k=0,2,4$ respectively, and $\Phi$ is the dilaton.
For later convenience, we set
\be
S= e^\Phi \tilde S~,~~~F= e^\Phi \tilde F~,~~~G= e^\Phi \tilde G~.
\ee
In addition to the Killing spinor equations the proof will also rely on the field equations and Bianchi identities (for relevant expressions in the conventions introduced above see \cite{Gran:2014rwa}), and the result will thus hold in general for all supersymmetric supergravity solutions.

It is convenient to introduce the following notation
\bea
&& \alpha^{IJ}_{B_1 \cdots B_k} \equiv B(\epsilon^I, \Gamma_{B_1  \cdots B_k}\epsilon^J) ~, \notag \\
&& \tau^{IJ}_{B_1 \cdots B_k} \equiv B(\epsilon^I, \Gamma_{B_1  \cdots B_k} \tilde\epsilon^J) ~,
\eea
where $\tilde\epsilon = \Gamma_{11}\epsilon$, the inner product $B(\epsilon^I, \epsilon^J) \equiv \langle  \Gamma_0 C* \epsilon^I, \epsilon^J \rangle$, where $C=\Gamma_{6789}$, is antisymmetric, i.e.~$B(\epsilon^I, \epsilon^J)=-B(\epsilon^J, \epsilon^I)$ and all $\Gamma$-matrices are anti-Hermitian with respect to this inner product, i.e.~$B(\Gamma_A \epsilon^I,\epsilon^J)=-B(\epsilon^I,\Gamma_A \epsilon^J)$.

Denoting $\alpha^{IJ}_{B_1 \cdots B_k} =\alpha^{IJ}_{(k)}$ and $\tau^{IJ}_{B_1 \cdots B_k} =\tau^{IJ}_{(k)}$ the bilinears have the symmetry properties
\bea
&&\alpha^{IJ}_{(k)} = \alpha^{JI}_{(k)}  \qquad  k=1,2,5 \notag ~ \\
&&\alpha^{IJ}_{(k)} = -\alpha^{JI}_{(k)} \quad ~ k=0,3,4
\eea
and
\bea
&&\tau^{IJ}_{(k)} = \tau^{JI}_{(k)}  \qquad  k=0,1,4,5 \notag ~ \\
&&\tau^{IJ}_{(k)} = -\tau^{JI}_{(k)} \quad  ~k=2,3 ~.
\eea

First we verify that there is a set of 1-form bi-linears whose associated vectors are Killing.  We write the gravitino KSE as
\bea
&&(\nabla_A + \Sigma_A)\epsilon = 0
\eea
where
\bea
&& \Sigma_A =  \tfrac{1}{8} H_{MP_1 P_2}\Gamma^{P_1 P_2}\Gamma_{11} +\tfrac{1}{8}  S\Gamma_M +\tfrac{1}{16} F_{P_1 P_2}\Gamma^{P_1 P_2}\Gamma_M \Gamma_{11}   \notag \\
&& \qquad\quad +\tfrac{1}{8\cdot 4!}G_{P_1 \cdots P_4}\Gamma^{P_1 \cdots P_4}\Gamma_M
 ~,
\eea
which we use to replace covariant derivatives with fluxes and $\Gamma$-matrixes. The 1-form bilinears associated with the Killing vectors are $\alpha^{IJ}_A e^A$, which we see by computing
\bea
\nabla_A \alpha^{IJ}_B &=& \nabla_A B(\epsilon^{I}, \Gamma_B \epsilon^{J}) \notag \\
&=& B(\nabla_A \epsilon^{I}, \Gamma_B  \epsilon^{J}) +B( \epsilon^{I}, \Gamma_B \nabla_A\epsilon^{J}) \notag \\
&=& -B(\Sigma_A \epsilon^{I}, \Gamma_B \epsilon^{J}) -B( \epsilon^{I}, \Gamma_B \Sigma_A\epsilon^{J}) \notag \\
&=& B(\Gamma_B \epsilon^{J},\Sigma_A \epsilon^{I}) -B( \epsilon^{I}, \Gamma_B  \Sigma_A\epsilon^{J})  \\
&=&  -B(\epsilon^{J},  \Gamma_B \Sigma_A \epsilon^{I}) -B( \epsilon^{I}, \Gamma_B  \Sigma_A\epsilon^{J}) \notag \\
&=& -2  B(\epsilon^{(I}, \Gamma_{B}\Sigma_A \epsilon^{J)})\notag \\
&=&\Big(\frac{1}{4} S \alpha^{IJ}_{AB} + \frac{1}{8}G_{A B}{}^{ C_1 C_2}\alpha^{IJ}_{C_1 C_2} + \frac{1}{96}G^{C_1 C_2 C_3 C_4} \alpha^{IJ}_{A B C_1 C_2 C_3 C_4} \notag \\
&& +\frac{1}{4} F_{A B} \tau^{IJ} - \frac{1}{2} H_{AB}{}^C \tau^{IJ}_C +\frac{1}{8} F^{C_1 C_2} \tau^{IJ}_{AB C_1 C_2}\Big) \notag ~.
\eea
Since the resulting expression is antisymmetric in its free indices we find that $\nabla_{(A} \alpha^{IJ}_{B)}=0 $ and hence the vectors associated with $\alpha^{IJ}_A e^A$ are Killing.

Note that the dilatino KSE (\ref{akse}) imply that
\bea
0  = B(\epsilon^{(I},{\cal A} \epsilon^{J)}) = \alpha^{IJ}_A \partial^A \Phi~,
\eea
and hence $i_K d\Phi=0$, where $K=\alpha^{IJ}_A e^A$ denotes the 1-forms associated with the Killing vectors with the $IJ$ indices suppressed. With this relation it follows that the Killing vectors preserve the dilaton:
\bea
{\cal L}_K \Phi := i_K d\Phi + d (i_K \Phi) =  0 ~,
\eea
since $i_K \Phi \equiv 0$.

To see that the 3-form flux $H$ is preserved we need to analyse the 1-form bi-linears which are not related to the Killing vectors, i.e. $\tau^{IJ}_A e^A$. As above, we find that
\bea
\nabla_{[A} \tau^{IJ}_{B]} &=& -2 B(\epsilon^{(I},\Gamma_{11} \Gamma_{[A} \Sigma_{B]} \epsilon^{J)}) \notag \\
&& = - \frac{1}{2} H_{AB}{}^C \alpha^{IJ}_C ~,
\label{3formeq}
\eea
or equivalently
\bea
d\tau_{(1)}^{IJ} = -i_K H ~,
\label{3formeq2}
\eea
where we have indicated the degree of the form $\tau$ and suppressed the indices labelling the Killing spinors.
By taking the exterior derivative of (\ref{3formeq2}), and using the Bianchi identity for $H$, i.e.~$dH=0$,
it follows that
\bea
{\cal L}_K H = 0
\eea
and hence the Killing vectors preserve also the $H$ flux.

We now turn to the 2-form flux $F$. Computing the (covariant) derivative of the scalar $\tau^{IJ}$, and making use of the gravitino KSE as above, we find
\bea
d\tau_{(0)}^{IJ} = i_K F + d\Phi \,\tau_{(0)}^{IJ} - S \,\tau_{(1)}^{IJ} ~.
\label{Fexpr}
\eea
Acting with another derivative on (\ref{Fexpr}), and re-substituting (\ref{Fexpr}) into the resulting expression, we obtain
\bea
0 = {\cal L}_K F - i_K (dF) -d\Phi \wedge i_K F - S d\tau_{(1)}^{IJ} = {\cal L}_K F ~,
\eea
where in the second step we have used  (\ref{3formeq2}) and the Bianchi identity for $F$, i.e.
\bea
0 = dF - d\Phi \wedge F -S H ~.
\label{FBI}
\eea
This shows that the $F$ flux is preserved.

For the $G$ flux a similar analysis can be preformed. Computing the covariant derivative of $\alpha_{(2)}^{IJ}$ leads to
\bea
d \alpha_{(2)}^{IJ} = i_K G + d\Phi \wedge \alpha_{(2)}^{IJ} + H \,\tau_{(0)}^{IJ} - F \wedge \tau^{IJ}_{(1)} ~.
\label{Gexpr}
\eea
Acting with an exterior derivative on (\ref{Gexpr}) and re-substituting (\ref{Gexpr}) into the resulting expression, and using (\ref{3formeq2}), (\ref{Fexpr}) and (\ref{FBI}), we obtain
\bea
0 = {\cal L}_K G - i_K (dG) - d\Phi \wedge i_K G + i_K F \wedge  H + F \wedge i_K H = {\cal L}_K G~,
\eea
where in the second step we have used the Bianchi identity for $G$, i.e.
\bea
0 = dG - d\Phi \wedge G - F \wedge H ~,
\eea
concluding the proof of the preservation of $G$.

Finally, ${\cal L}_K S=0$ follows from the constancy of the Romans mass parameter $\tilde S$, which completes the proof.
For the computations in this appendix the Mathematica package GAMMA \cite{Gran:2001yh} has been used.

\setcounter{subsection}{0}
\setcounter{equation}{0}

\appendix{AdS Superalgebra}

Here we collect some of the key formulae that are needed to determine the superalgebras of AdS backgrounds. The proof  we have presented relies on
the observation that the commutators of the superalgebra can be computed explicitly when the generators are associated
with symmetries of the AdS subspace of the background.  For this we give the Killing vectors of AdS subspace and
their commutators as well as their action on the Killing spinors.

\subsection{Isometries of AdS}

The associated 1-forms of the Killing vectors  along the AdS subspace of   AdS$_k \times_w M^k$, $n \geq 3$  equipped with the metric (\ref{metr})
are

\begin{gather}
 \lambda^+ = A^2 e^{2 z / \ell} \bbe^+~, \qquad \lambda^- =  \bbe^-~, \qquad \lambda^a = A e^{ z / \ell} \bbe^a~,
 \nonumber \\
 \lambda^z = A \bbe^z - \ell^{-1} \hat{r} \lambda^+ - \ell^{-1} u \lambda^- - \ell^{-1} x_a \lambda^a, \qquad M^{+ -} = \hat{r} \lambda^+ - u \lambda^-~,
 \nonumber \\
 M^{+ a} = x^a \lambda^+ - u \lambda^a~, \qquad M^{- a} = x^a \lambda^- - \hat{r} \lambda^a~, \qquad M^{a b} = x^b \lambda^a - x^a \lambda^b~,
 \nonumber \\
 M^{z +} = p \lambda^+ + u \lambda^z~, \qquad M^{z -} = p \lambda^- + \hat{r} \lambda^z~, \qquad M^{z a} = p \lambda^a + x^a \lambda^z~,
\end{gather}
where $p = \frac{1}{2} ( 2 \ell^{-1} u \hat{r} + \ell e^{-2 z / \ell} - \ell + \ell^{-1} \mathbf{x}^2 ) $ and $\hat{r} = r A^{-2} e^{-2 z / \ell}$.
Note that $\ell$ is the radius of AdS$_k$, the warp factor, $A$, depends only on the coordinates $y^i$ of $M^k$ while $u,r,z, x^a$ for $a=1,\dots, n-3$ are the coordinates of AdS$_k$ subspace, and $g_{ij}$ is the metric on $M^k$.

The associated Killing vectors are
\begin{gather}
 \kappa^+ = A^2 e^{2z / \ell} \partial_r~, \qquad \kappa^- = \partial_u~, \qquad \kappa^a = \partial_a~,
 \nonumber \\
 \kappa^z = \partial_z + \ell^{-1} r \partial_r - \ell^{-1} u \partial_u - \ell^{-1} x^a \partial_a~, \qquad N^{+ -} = r \partial_r - u \partial_u~,
 \nonumber \\
 \qquad N^{+ a} = A^2 e^{2z / \ell} x^a \partial_r - u \partial_a~, \qquad N^{- a} = x^a \partial_u - A^{-2} e^{-2z / \ell} r \partial_a~, \qquad N^{a b} = x^b \partial_a - x^a \partial_b~,
 \nonumber \\
 N^{z +} = A^2 e^{2z / \ell} p \partial_r + u \kappa^z~, \qquad N^{z -} = p \partial_u + A^{-2} e^{-2z / \ell} r \kappa^z~, \qquad N^{z a} = p \partial_a + x^a \kappa^z~.
\end{gather}

Moreover, the commutators of these Killing vectors are
\begin{gather}
 [ \lambda^\mu, \lambda^\nu ] = \ell^{-1} f^{\mu \nu}{}_\kappa \lambda^\kappa~, \quad [ \lambda^\mu, M^{\nu \sigma} ] = -2 \eta^{\mu [ \nu} \lambda^{\sigma ]} + \ell^{-1} f^{\mu \nu \sigma}{}_{\kappa_1 \kappa_2} M^{\kappa_1 \kappa_2}~,
 \nonumber \\
 [ M^{\mu \nu}, M^{\sigma \tau} ] = 2 ( \eta^{\mu [ \tau} M^{\sigma ] \nu} - \eta^{\nu [ \tau} M^{\sigma ] \mu} ) ,
\end{gather}
where
\begin{gather}
 f^{z \pm}{}_\pm = 1~, \quad f^{z a}{}_b = \delta^a_b~, \quad f^{\pm \mp z}{}_{[+ -]} = \pm \tfrac{1}{2}~, \quad f^{a \pm z}{}_{[\pm b]} =- \tfrac{1}{2} \delta^a_b
 \nonumber \\
 f^{a z b}{}_{c d} = -\delta^a_{[c} \delta^b_{d]}~, \quad f^{z \pm z}{}_{[\pm z]} = -\tfrac{1}{2}~, \quad f^{z z a}{}_{[z b]} = -\tfrac{1}{2} \delta^a_b~, \quad f^{\pm za}{}_{\pm b}=-\tfrac{1}{2} \delta^a{}_b.
\end{gather}
The Lie algebra is isomorphic to $\mathfrak{so}(n-1, 2)$ as expected.

\subsection{Spinorial Lie Derivatives along AdS}

The commutators of the even generators associated with the isometries of AdS and odd generators associated with the Killing spinors of the superalgebra can also be explicitly found
via the spinorial derivative (\ref{spinder}) of the isometries of AdS on the Killing spinors of the backgrounds. In particular, one has
\begin{align}
 \mathcal{L}_{\lambda^+} \epsilon &= A^2 e^{2 z / \ell} \partial_r \epsilon~, \quad \mathcal{L}_{\lambda^-} \epsilon = \partial_u \epsilon~, \quad \mathcal{L}_{\lambda^a} \epsilon = \partial_a \epsilon~,
 \nonumber \\
 \mathcal{L}_{\lambda^z} \epsilon &= ( \partial_z + \ell^{-1} r \partial_r - \ell^{-1} u \partial_u - \ell^{-1} x^a \partial_a ) \epsilon + \frac{1}{2} \ell^{-1} \Gamma_{+ -} \epsilon
 \nonumber \\
 \mathcal{L}_{M^{+ -}} \epsilon &= ( r \partial_r - u \partial_u ) \epsilon + \frac{1}{2} \Gamma_{+ -} \epsilon
 \nonumber \\
 \mathcal{L}_{M^{+ a}} \epsilon &= ( x_a A^2 e^{2 z / \ell} \partial_r - u \partial_a ) \epsilon - \frac{1}{2} A e^{z / \ell} \Gamma_{- a} \epsilon
 \nonumber \\
 \mathcal{L}_{M^{- a}} \epsilon &= ( x_a \partial_u - \hat{r} \partial_a ) \epsilon - \frac{1}{2} A^{-1} e^{-z / \ell} \Gamma_{+ a} \epsilon
 \nonumber \\
 \mathcal{L}_{M^{a b}} \epsilon &= (x_b \partial_a - x_a \partial_b) \epsilon - \frac{1}{2} \Gamma_{a b} \epsilon
 \nonumber \\
 \mathcal{L}_{M^{z +}} \epsilon &= \left[ ( p + \ell^{-1} u \hat{r} ) A^2 e^{2 z / \ell} \partial_r + u \partial_z - \ell^{-1} u^2 \partial_u - \ell^{-1} u x^a \partial_a \right] \epsilon
 \nonumber \\
 & \qquad + ( \frac{1}{2} A \Gamma_{- z} + \ell^{-1} u \Gamma_{+ -} - \frac{1}{2} \ell^{-1} A e^{z / \ell} x^a \Gamma_{- a} ) \epsilon
 \nonumber \\
 \mathcal{L}_{M^{z -}} \epsilon &= \left[ ( p - \ell^{-1} u \hat{r} ) \partial_u + \hat{r} \partial_z + \ell^{-1} r \hat{r} \partial_r - \ell^{-1} \hat{r} x^a \partial_a \right] \epsilon
 \nonumber \\
 & \qquad + ( \frac{1}{2} A^{-1} e^{-2 z / \ell} \Gamma_{+ z} - \frac{1}{2} \ell^{-1} A^{-1} e^{-z / \ell} x^a \Gamma_{+ a} ) \epsilon
 \nonumber \\
 \mathcal{L}_{M^{z a}} \epsilon &= \left[ (p \delta_a^b - \ell^{-1} x_a x^b) \partial_b + x_a \partial_z + \ell^{-1} r x_a \partial_r - \ell^{-1} u x_a \partial_u \right] \epsilon
 \nonumber \\
 & \qquad + ( \frac{1}{2} \ell^{-1} x_a \Gamma_{+ -} + \frac{1}{2} \ell^{-1} u A^{-1} e^{-z / \ell} \Gamma_{+ a}
 \nonumber \\
 & \qquad + \frac{1}{2} \ell^{-1} r A^{-1} e^{-z / \ell} \Gamma_{- a} - \frac{1}{2} e^{-z / \ell} \Gamma_{z a} - \frac{1}{2} \ell^{-1} x^b \Gamma_{a b} ) \epsilon~.
\end{align}
These spinorial Lie derivatives determine all such commutators as the dependence of the Killing spinors on the AdS coordinates is known.  The super-Jacobi identities are then used to restrict the remaining  commutators of the even generators
associated with the isometries of the internal space with odd generators of the superalgebra.

\setcounter{subsection}{0}
\setcounter{equation}{0}

\appendix{Commutators} \label{allbilinearsxx}

\subsection{1-form bilinears}

The linearly independent Killing spinors of AdS$_k$, $n\geq 3$, backgrounds expressed in terms  of $\sigma_\pm$ and $\tau_\pm$  are
\bea
\epsilon_1(\sigma_+)&=&\sigma_+~,~~~\epsilon_2(\sigma_-)=\sigma_--\ell^{-1} e^{{z\over\ell}} x^a \Gamma_{az} \sigma_--\ell^{-1} A^{-1} u \Gamma_{+z} \sigma_-
\cr
\epsilon_3(\tau_+)&=&e^{-{z\over\ell}}\tau_+-\ell^{-1} A^{-1} r e^{-{z\over\ell}} \Gamma_{-z} \tau_+-\ell^{-1} x^a \Gamma_{az} \tau_+~,~~~\epsilon_4(\tau_-)=e^{{z\over \ell}} \tau_-~.
\label{kssigma}
\eea
For $n=3$, the terms proportional to $x^a$ do not occur and they should be set to zero.

The anti-commutator of the odd generators can be founds from the 1-form bilinears. These are
\bea
K(\epsilon_1, \epsilon_1)&=&2 \parallel\sigma_+\parallel^2 \,\bbe^-~,~~~
\cr
K(\epsilon_2, \epsilon_2)&=&-2 (1+\ell^{-2} x^2 e^{{2z\over\ell}}) \parallel\sigma_-\parallel^2 \,\bbe^++4 \ell^{-2} A^{-2} u^2 \parallel\sigma_-\parallel^2 \,\bbe^-
\cr &&~~~-4 \ell^{-1} A^{-1} u \parallel\sigma_-\parallel^2 \,\bbe^z+ 4 \ell^{-2} A^{-1} e^{{z\over\ell}} u x_a \parallel\sigma_-\parallel^2 \,\bbe^a
\cr
K(\epsilon_3, \epsilon_3)&=&-4\ell^{-2}  A^{-2} r^2 e^{-{2z\over\ell}} \parallel \tau_+\parallel^2 \,\bbe^++ 2 (e^{-{2z\over\ell}}+ \ell^{-2} x^2) \parallel \tau_+\parallel^2 \,\bbe^-\cr
&&~~~~+4 \ell^{-1} A^{-1} r e^{-{2z\over\ell}} \parallel \tau_+\parallel^2 \,\bbe^z-4 \ell^{-2} A^{-1} r e^{-{z\over\ell}} x_a \parallel \tau_+\parallel^2 \,\bbe^a
\cr
K(\epsilon_4, \epsilon_4)&=&-2 e^{{2z\over\ell}}\parallel\tau_-\parallel^2 \,\bbe^+~,~~
\cr
K(\epsilon_2, \epsilon_1)&=&-2 \ell^{-1} A^{-1} u \langle \Gamma_{+z}\sigma_-, \sigma_+\rangle \,\bbe^-+[\langle \Gamma_{+z}\sigma_-, \sigma_+\rangle-\ell^{-1} e^{{z\over\ell}} x^a
\langle \Gamma_{+}\sigma_-, \Gamma_a\sigma_+\rangle] \,\bbe^z
\cr &&~~~~-[ \langle \Gamma_{+}\sigma_-, \Gamma_c\sigma_+\rangle+\ell^{-1} e^{{z\over\ell}} x^a \langle \Gamma_{+z}\sigma_-, \Gamma_a\Gamma_c \sigma_+\rangle] \,\bbe^c
\cr &&~~~~-[ \langle \Gamma_{+}\sigma_-, \Gamma_i\sigma_+\rangle+\ell^{-1} e^{{z\over\ell}} x^a \langle \Gamma_{+z}\sigma_-, \Gamma_a\Gamma_i \sigma_+\rangle] \,\bbe^i
\cr
K(\epsilon_3, \epsilon_1)&=& 2 (e^{-{z\over\ell}}\langle \tau_+, \sigma_+\rangle -\ell^{-1} x^a \langle \Gamma_{a z} \tau_+, \sigma_+\rangle) \,\bbe^-
\cr &&~~~~~+2 \ell^{-1} A^{-1} r e^{-{z\over\ell}}
\langle \tau_+, \sigma_+\rangle \,\bbe^z+2 \ell^{-1} A^{-1} r e^{-{z\over\ell}}
\langle \tau_+, \Gamma_{za}\sigma_+\rangle \,\bbe^a
\cr&&~~~~~+2 \ell^{-1} A^{-1} r e^{-{z\over\ell}}
\langle \tau_+,\Gamma_{zi} \sigma_+\rangle \,\bbe^i
\cr
K(\epsilon_4, \epsilon_1)&=&- e^{{z\over\ell}} \langle \Gamma_+\tau_-,\Gamma_{z} \sigma_+\rangle \,\bbe^z-e^{{z\over\ell}} \langle \Gamma_+\tau_-,\Gamma_{a} \sigma_+\rangle \,\bbe^a-e^{{z\over\ell}} \langle \Gamma_+\tau_-,\Gamma_{i} \sigma_+\rangle \,\bbe^i
\cr
K(\epsilon_2, \epsilon_3)&=&2 \ell^{-1} A^{-1} r [- e^{-{z\over\ell}}\langle\Gamma_{+z}\sigma_-, \tau_+\rangle+ \ell^{-1} x^a \langle \Gamma_+\sigma_-, \Gamma_a \tau_+\rangle] \,\bbe^+
\cr &&
+2 \ell^{-1} A^{-1} u [-e^{-{z\over\ell}}\langle\Gamma_{+z}\sigma_-, \tau_+\rangle+ \ell^{-1} x^a \langle \Gamma_+\sigma_-, \Gamma_a \tau_+\rangle] \,\bbe^-
\cr &&
+[e^{-{z\over\ell}}\langle\Gamma_{+z}\sigma_-, \tau_+\rangle-2\ell^{-1} x^a \langle \Gamma_+\sigma_-, \Gamma_a \tau_+\rangle-\ell^{-2} e^{{z\over\ell}} x^2\langle\Gamma_{+z}\sigma_-, \tau_+\rangle
\cr && -2 \ell^{-2}  A^{-2} u r e^{-{z\over\ell}} \langle\Gamma_{+z}\sigma_-, \tau_+\rangle] \,\bbe^z
\cr &&
+[-e^{-{z\over\ell}}\langle\Gamma_{+}\sigma_-, \Gamma_c\tau_+\rangle- 2 \ell^{-1} x_c \langle\Gamma_{+z}\sigma_-, \tau_+\rangle+ \ell^{-2} e^{{z\over\ell}} x^a x^b
\langle\Gamma_{+}\sigma_-, \Gamma_a \Gamma_c \Gamma_b \tau_+\rangle
\cr &&
- 2 \ell^{-2} A^{-2} u r e^{-{z\over\ell}}  \langle\Gamma_{+}\sigma_-, \Gamma_c\tau_+\rangle] \,\bbe^c
\cr &&
- [e^{-{z\over\ell}}+\ell^{-2}e^{{z\over\ell}} x^2+2 \ell^{-2} A^{-2} ur e^{-{z\over\ell}}] \langle\Gamma_{+}\sigma_-, \Gamma_i\tau_+\rangle \,\bbe^i
\cr
K(\epsilon_2, \epsilon_4)&=&-2 [ e^{{z\over\ell}}\langle\sigma_-, \tau_-\rangle-\ell^{-1} e^{{2z\over\ell}} x^a \langle\sigma_-,\Gamma_{za} \tau_-\rangle] \,\bbe^+-2 \ell^{-1} A^{-1} u e^{{z\over\ell}}
\langle\sigma_-, \tau_-\rangle \,\bbe^z
\cr &&
-2 \ell^{-1} A^{-1} u e^{{z\over\ell}}\langle\sigma_-,\Gamma_{za} \tau_-\rangle \,\bbe^a-2 \ell^{-1} A^{-1} u e^{{z\over\ell}}\langle\sigma_-,\Gamma_{zi} \tau_-\rangle \,\bbe^i
\cr
K(\epsilon_3, \epsilon_4)&=&-2 \ell^{-1} r A^{-1}\langle\tau_+,\Gamma_{+z} \tau_-\rangle \,\bbe^++[\langle\tau_+,\Gamma_{+z} \tau_-\rangle+ \ell^{-1} x^a e^{{z\over\ell}}
\langle\tau_+,\Gamma_{+a} \tau_-\rangle] \,\bbe^z
\cr
&&
+[\langle\tau_+,\Gamma_{+c} \tau_-\rangle-\ell^{-1} x^a e^{{z\over\ell}}\langle\tau_+,\Gamma_a \Gamma_c\Gamma_{+z} \tau_-\rangle] \,\bbe^c
\cr &&
+ [\langle\tau_+,\Gamma_{+i} \tau_-\rangle-\ell^{-1} x^a
 e^{{z\over\ell}} \langle\tau_+,\Gamma_a \Gamma_i\Gamma_{+z} \tau_-\rangle] \,\bbe^i~.
\eea
Note that bilinears of  different Killing spinors $\epsilon^r$ are given as above after replacing $\parallel\sigma_+\parallel^2$ with the inner product
$\langle \sigma_+^r, \sigma_+^s\rangle$ and similarly for the rest of the $\sigma_\pm$ and $\tau_\pm$ spinors.

After imposing the conditions that arise from the global considerations, the form of the bilinears simplifies to

\bea
K(\epsilon_1, \epsilon_1)&=&2 \parallel\sigma_+\parallel^2 \,\bbe^-  ~,~~~
\cr
K(\epsilon_2, \epsilon_2)&=&-2 (1+\ell^{-2} x^2 e^{{2z\over\ell}}) \parallel\sigma_-\parallel^2 \,\bbe^++4 \ell^{-2} A^{-2} u^2 \parallel\sigma_-\parallel^2 \,\bbe^-
\cr &&~~~-4 \ell^{-1} A^{-1} u \parallel\sigma_-\parallel^2 \,\bbe^z+ 4 \ell^{-2} A^{-1} e^{{z\over\ell}} u x_a \parallel\sigma_-\parallel^2 \,\bbe^a
\cr
K(\epsilon_3, \epsilon_3)&=&-4\ell^{-2}  A^{-2} r^2 e^{-{2z\over\ell}} \parallel \tau_+\parallel^2 \,\bbe^++ 2 (e^{-{2z\over\ell}}+ \ell^{-2} x^2) \parallel \tau_+\parallel^2 \,\bbe^-\cr
&&~~~~+4 \ell^{-1} A^{-1} r e^{-{2z\over\ell}} \parallel \tau_+\parallel^2 \,\bbe^z-4 \ell^{-2} A^{-1} r e^{-{z\over\ell}} x_a \parallel \tau_+\parallel^2 \,\bbe^a
\cr
K(\epsilon_4, \epsilon_4)&=&-2 e^{{2z\over\ell}}\parallel\tau_-\parallel^2 \,\bbe^+
\cr
K(\epsilon_2, \epsilon_1)&=&-2 \ell^{-1} A^{-1} u \langle \Gamma_{+z}\sigma_-, \sigma_+\rangle \,\bbe^-+\langle \Gamma_{+z}\sigma_-, \sigma_+\rangle \,\bbe^z
\cr &&~~~~-\ell^{-1} e^{{z\over\ell}} x^a \langle \Gamma_{+z}\sigma_-, \Gamma_a\Gamma_c \sigma_+\rangle \,\bbe^c
- \langle \Gamma_{+}\sigma_-, \Gamma_i\sigma_+\rangle \,\bbe^i
\cr
K(\epsilon_3, \epsilon_1)&=&  -2\ell^{-1} x^a \langle \Gamma_{a z} \tau_+, \sigma_+\rangle \,\bbe^-
+2 \ell^{-1} A^{-1} r e^{-{z\over\ell}}
\langle \tau_+, \Gamma_{za}\sigma_+\rangle \,\bbe^a
\cr
K(\epsilon_4, \epsilon_1)&=&-e^{{z\over\ell}} \langle \Gamma_+\tau_-,\Gamma_{a} \sigma_+\rangle \,\bbe^a
\cr
K(\epsilon_2, \epsilon_3)&=&2 \ell^{-2} A^{-1} r  x^a \langle \Gamma_+\sigma_-, \Gamma_a \tau_+\rangle \,\bbe^+
+2 \ell^{-2} A^{-1} u  x^a \langle \Gamma_+\sigma_-, \Gamma_a \tau_+\rangle  \,\bbe^-
\cr &&
-2\ell^{-1} x^a \langle \Gamma_+\sigma_-, \Gamma_a \tau_+\rangle \,\bbe^z
+[-e^{-{z\over\ell}}\langle\Gamma_{+}\sigma_-, \Gamma_c\tau_+\rangle+ \ell^{-2} e^{{z\over\ell}} x^a x^b
\langle\Gamma_{+}\sigma_-, \Gamma_a \Gamma_c \Gamma_b \tau_+\rangle
\cr &&
- 2 \ell^{-2} A^{-2} u r e^{-{z\over\ell}}  \langle\Gamma_{+}\sigma_-, \Gamma_c\tau_+\rangle] \,\bbe^c
\cr
K(\epsilon_2, \epsilon_4)&=&2 \ell^{-1} e^{{2z\over\ell}} x^a \langle\sigma_-,\Gamma_{za} \tau_-\rangle \,\bbe^+
-2 \ell^{-1} A^{-1} u e^{{z\over\ell}}\langle\sigma_-,\Gamma_{za} \tau_-\rangle \,\bbe^a
\cr
K(\epsilon_3, \epsilon_4)&=&-2 \ell^{-1} r A^{-1}\langle\tau_+,\Gamma_{+z} \tau_-\rangle \,\bbe^++\langle\tau_+,\Gamma_{+z} \tau_-\rangle  \,\bbe^z
-\ell^{-1} x^a e^{{z\over\ell}}\langle\tau_+,\Gamma_a \Gamma_c\Gamma_{+z} \tau_-\rangle \,\bbe^c
\cr &&
+ \langle\tau_+,\Gamma_{+i} \tau_-\rangle \,\bbe^i
\eea
For $n>2$, we know that $\sigma_-=A\Gamma_{-z} \sigma'_+$ and $\tau_-=A \Gamma_{-z} \tau'_+$.  Using these conditions, the bilinears can be written in the basis of Killing
 vectors of AdS given in appendix B as
\bea
K(\epsilon_1, \epsilon_1)&=&2 \parallel\sigma_+\parallel^2 \,\bbe^-=2 \parallel\sigma_+\parallel^2 \lambda^-  ~,~~~
\cr
K(\epsilon_2, \epsilon_2)&=&-2 (1+\ell^{-2} x^2 e^{{2z\over\ell}}) \parallel\sigma_-\parallel^2 \,\bbe^++4 \ell^{-2} A^{-2} u^2 \parallel\sigma_-\parallel^2 \,\bbe^-
\cr &&~~~-4 \ell^{-1} A^{-1} u \parallel\sigma_-\parallel^2 \,\bbe^z+ 4 \ell^{-2} A^{-1} e^{{z\over\ell}} u x_a \parallel\sigma_-\parallel^2 \,\bbe^a
\cr &=&-2 (4\ell^{-1} M^{z+}+2 \lambda^+)  \parallel\sigma_+'\parallel^2
\cr
K(\epsilon_3, \epsilon_3)&=&-4\ell^{-2}  A^{-2} r^2 e^{-{2z\over\ell}} \parallel \tau_+\parallel^2 \,\bbe^++ 2 (e^{-{2z\over\ell}}+ \ell^{-2} x^2) \parallel \tau_+\parallel^2 \,\bbe^-\cr
&&~~~~+4 \ell^{-1} A^{-1} r e^{-{2z\over\ell}} \parallel \tau_+\parallel^2 \,\bbe^z-4 \ell^{-2} A^{-1} r e^{-{z\over\ell}} x_a \parallel \tau_+\parallel^2 \,\bbe^a
\cr
&=& (4\ell^{-1} M^{z-}+2 \lambda^-) \parallel \tau_+\parallel^2
\cr
K(\epsilon_4, \epsilon_4)&=&-2 e^{{2z\over\ell}}\parallel\tau_-\parallel^2 \,\bbe^+=-4 \parallel\tau_+'\parallel^2 \lambda^+
\cr
K(\epsilon_2, \epsilon_1)&=&4 \ell^{-1}  u \langle \sigma'_+, \sigma_+\rangle \,\bbe^--2A\langle  \sigma'_+, \sigma_+\rangle \,\bbe^z
+2A\ell^{-1} e^{{z\over\ell}} x^a \langle \sigma'_+, \Gamma_a\Gamma_c \sigma_+\rangle \,\bbe^c
- 2 A\langle \sigma'_+, \Gamma_{zi}\sigma_+\rangle \,\bbe^i
\cr &=& -2 (\ell^{-1} M^{+-}+\lambda^z) \langle \sigma'_+, \sigma_+\rangle-\ell^{-1}  M^{ab} \langle \sigma'_+, \Gamma_{ab}\sigma_+\rangle - 2 A\langle \sigma'_+, \Gamma_{zi}\sigma_+\rangle \,\bbe^i
\cr
K(\epsilon_3, \epsilon_1)&=&  -2\ell^{-1} x^a \langle \Gamma_{a z} \tau_+, \sigma_+\rangle \,\bbe^-
+2 \ell^{-1} A^{-1} r e^{-{z\over\ell}}
\langle \tau_+, \Gamma_{za}\sigma_+\rangle \,\bbe^a
\cr
&=& -2\ell^{-1} M^{-a} \langle \Gamma_{a z} \tau_+, \sigma_+\rangle
\cr
K(\epsilon_4, \epsilon_1)&=&-2 A e^{{z\over\ell}} \langle \tau'_+,\Gamma_{za} \sigma_+\rangle \,\bbe^a= -2 \lambda^a \langle \tau'_+,\Gamma_{za} \sigma_+\rangle
\cr
K(\epsilon_2, \epsilon_3)&=&4 \ell^{-2}  r  x^a \langle \sigma_+', \Gamma_{za} \tau_+\rangle \,\bbe^+
+4 \ell^{-2} u  x^a \langle \sigma_+', \Gamma_{za} \tau_+\rangle  \,\bbe^-
\cr &&
-4 A\ell^{-1} x^a \langle \sigma_+', \Gamma_{za} \tau_+\rangle \,\bbe^z
+[-2 A e^{-{z\over\ell}}\langle\sigma_+', \Gamma_{zc}\tau_+\rangle+ 2 A \ell^{-2} e^{{z\over\ell}} x^a x^b
\langle\sigma'_+, \Gamma_{za} \Gamma_c \Gamma_b \tau_+\rangle
\cr &&
- 4 \ell^{-2} A^{-1} u r e^{-{z\over\ell}}  \langle\sigma_+', \Gamma_{zc}\tau_+\rangle] \,\bbe^c
\cr
&=& -(4\ell^{-1} M^{za} +2 \lambda^a) \langle \sigma_+', \Gamma_{za} \tau_+\rangle
\cr
K(\epsilon_2, \epsilon_4)&=&-4A^2 \ell^{-1} e^{{2z\over\ell}} x^a \langle\sigma_+',\Gamma_{za} \tau_+'\rangle \,\bbe^+
+4 A \ell^{-1}  u e^{{z\over\ell}}\langle\sigma_+',\Gamma_{za} \tau_+'\rangle \,\bbe^a
\cr
&=&-4\ell^{-1} M^{+a} \langle\sigma_+',\Gamma_{za} \tau_+'\rangle
\cr
K(\epsilon_3, \epsilon_4)&=&4 \ell^{-1} r \langle\tau_+, \tau_+'\rangle \,\bbe^+-2A \langle\tau_+, \tau_+'\rangle  \,\bbe^z
+2A\ell^{-1} x^a e^{{z\over\ell}}\langle\tau_+,\Gamma_a \Gamma_c \tau_+'\rangle \,\bbe^c
\cr &&
+ 2A \langle\tau_+,\Gamma_{zi} \tau_+'\rangle \,\bbe^i
\cr
&=& 2 (-\lambda^z+\ell^{-1} M^{+-}) \langle\tau_+, \tau_+'\rangle-  \ell^{-1} M^{ab} \langle\tau_+,\Gamma_{ab}  \tau_+'\rangle + 2A \langle\tau_+,\Gamma_{zi} \tau_+'\rangle \,\bbe^i~.
\eea
From the above expressions it is straightforward to read the anti-commutator of any two odd generators of the superalgebra.
Of course, for $n=3$ one should neglect all the terms which involve isometries that carry a  $\bbe^a$ frame index.

\subsection{Spinorial Lie Derivative}

The computation of the commutator of even and odd generators requires the evaluation of the spinorial Lie derivative of the Killing
spinors along the isometries of AdS. In particular, the non-vanishing spinorial Lie derivatives of  $\epsilon_1(\sigma_+)$
along the isometries of AdS are as follows
\bea
&&{\cal L}_{\lambda^z}\epsilon_1(\sigma_+)={1\over2}\ell^{-1} \epsilon_1(\sigma_+)~,~~~{\cal L}_{M^{+-}}\epsilon_1(\sigma_+)={1\over2} \epsilon_1(\sigma_+)~,~~~
\cr
&&{\cal L}_{M^{+a}}\epsilon_1(\sigma_+)=-{1\over2}\epsilon_4(A \Gamma_{-a}\sigma_+)~,~~~{\cal L}_{M^{ab}}\epsilon_1(\sigma_+)=-{1\over2} \epsilon_1(\Gamma_{ab}\sigma_+)~,~~~
\cr
&&{\cal L}_{M^{z+}}\epsilon_1(\sigma_+)={1\over2} \epsilon_2(A\Gamma_{-z} \sigma_+)~,~~~{\cal L}_{M^{za}}\epsilon_1(\sigma_+)=-{1\over2} \epsilon_3(\Gamma_{za}\sigma_+) \ .
\eea
The  non-vanishing spinorial Lie derivatives of  $\epsilon_2(\sigma_-)$ are
\bea
&&{\cal L}_{\lambda^-}\epsilon_2(\sigma_-)=-\ell^{-1} \epsilon_1(A^{-1} \Gamma_{+z}\sigma_-)~,~~~{\cal L}_{\lambda^a}\epsilon_2(\sigma_-)=-\ell^{-1} \epsilon_4(\Gamma_{az}\sigma_-)~,
\cr
&&
{\cal L}_{\lambda^z}\epsilon_2(\sigma_-)=-{1\over2} \ell^{-1} \epsilon_2(\sigma_-)~,~~~{\cal L}_{M^{+-}}\epsilon_2(\sigma_-)=-{1\over2} \epsilon_2(\sigma_-)~,
\cr
&&
{\cal L}_{M^{ab}}\epsilon_2(\sigma_-)=-{1\over2} \epsilon_2(\Gamma_{ab}\sigma_-)~,~~~{\cal L}_{M^{-a}}\epsilon_2(\sigma_-)=-{1\over2} \epsilon_3(A^{-1}\Gamma_{+a} \sigma_-)~,
\cr
&&
{\cal L}_{M^{z-}}\epsilon_2(\sigma_-)={1\over2} \epsilon_1(A^{-1} \Gamma_{+z}\sigma_-)~,~~~{\cal L}_{M^{za}}\epsilon_2(\sigma_-)={1\over2} \epsilon_4(\Gamma_{az}\sigma_-) \ .
\eea
Similarly the non-vanishing spinorial Lie derivatives of$\epsilon_3(\tau_+)$ are
\bea
&&{\cal L}_{\lambda^+}\epsilon_3(\tau_+)=-\ell^{-1} \epsilon_4 (A\Gamma_{-z} \tau_+)~,~~~{\cal L}_{\lambda^a}\epsilon_3(\tau_+)=-\ell^{-1} \epsilon_1(\Gamma_{az} \tau_+)~,
\cr
&&
{\cal L}_{\lambda^z}\epsilon_3(\tau_+)=-{1\over2} \ell^{-1} \epsilon_3 (\tau_+)~,~~~{\cal L}_{M^{+-}}\epsilon_3(\tau_+)={1\over2} \epsilon_3(\tau_+)~,~~
\cr
&&
{\cal L}_{M^{+a}}\epsilon_3(\tau_+)=-{1\over2} \epsilon_2(A\Gamma_{-a} \tau_+)~,~~~{\cal L}_{M^{ab}}\epsilon_3(\tau_+)=-{1\over2} \epsilon_3(\Gamma_{ab} \tau_+)~,
\cr
&&
{\cal L}_{M^{z+}}\epsilon_3(\tau_+)={1\over2} \epsilon_4(A\Gamma_{-z} \tau_+)~,~~~{\cal L}_{M^{za}}\epsilon_3(\tau_+)={1\over2}\epsilon_1(\Gamma_{az} \tau_+)~,
\eea
and those of  $\epsilon_4(\tau_-)$ are
\bea
&&{\cal L}_{\lambda^z}\epsilon_4(\tau_-)={1\over2} \ell^{-1} \epsilon_4(\tau_-)~,~~~{\cal L}_{M^{+-}}\epsilon_4(\tau_-)=-{1\over2} \epsilon_4(\tau_-)~,
\cr
&&
{\cal L}_{M^{-a}}\epsilon_4(\tau_-)=-{1\over2} \epsilon_1(A^{-1} \Gamma_{+a}\tau_-)~,~~~{\cal L}_{M^{ab}}\epsilon_4(\tau_-)=-{1\over2} \epsilon_4(\Gamma_{ab}\tau_-)~,
\cr
&&
{\cal L}_{M^{z-}}\epsilon_4(\tau_-)={1\over2} \epsilon_3(A^{-1} \Gamma_{+z} \tau_-)~,~~~{\cal L}_{M^{za}}\epsilon_4(\tau_-)=-{1\over2} \epsilon_2(\Gamma_{za}\tau_-) \ .
\eea

\setcounter{subsection}{0}
\setcounter{equation}{0}

\appendix{AdS$_3$ KSAs for $N<14$}

\subsection{$N=N_\sigma=10$}

The 4-form $\alpha$ is dual to a 1-form. Because of this, the $SO(5)$ automorphisms of the KSA can be used to choose the 1-form to lie in the 5-th direction, i.e.~the only non-vanishing
component of $\alpha$ is $\alpha_{1234}$.
Next  consider the commutator
\bea
[\tilde V_{12}, \tilde V_{35}]~.
\eea
Using
\bea
\tilde V_{rs}={1\over2}\epsilon^{AB} \{Q_{Ar}, Q_{Bs}\}
\label{vqq}
\eea
and  after expressing first  $\tilde V_{12}$ and then $\tilde V_{35}$ in terms of $Q$'s, an application of the super-Jacobi identities reveals that
\bea
\alpha_{1234} \tilde V_{45}=0~.
\eea
Taking the commutator with $Q_{A4}$ and using the fact that all $Q$'s are linearly independent, we find that $\alpha_{1234}=0$ and so $\alpha=0$.  As a result the KSA is $\mathfrak{osp}(5,2)$.
This is in agreement with the general analysis presented in section 3 for the AdS$_3$ backgrounds.

\subsection{$N = N_\sigma = 12$}

The 4-form $\alpha$ is dual to a 2-form. Using  $SO(6)$ automorphisms of the KSA, ${}^*\alpha$  can be brought into the canonical form
\bea
{}^*\alpha= \lambda e^1\wedge e^2+ \lambda_2 e^3\wedge e^4+\lambda_3 e^5\wedge e^6
\eea
for some constants $\lambda_1, \lambda_2$ and $\lambda_3$.

It remains to determine these constants.  For this consider the commutator
\bea
[\tilde V_{12}, \tilde V_{35}]~.
\eea
After using (\ref{vqq}) to express first $\tilde V_{12}$ and then $\tilde V_{35}$ in terms of the $Q$'s and applying the super-Jacobi identity in both cases, we find that
\bea
-\lambda_3 \tilde V_{54}+\lambda_2 \tilde V_{36}=0~.
\eea
The commutators of the above expression with $Q_{A5}$ and $Q_{A3}$ reveal that
\bea
\lambda_3-\lambda_2 \lambda_1=0~,~~~\lambda_2-\lambda_3 \lambda_1=0~.
\label{x1con12}
\eea
A similar calculation involving the commutator $[\tilde V_{34}, \tilde V_{15}]$ leads to
\bea
\lambda_1-\lambda_3 \lambda_2=0~.
\label{x2con12}
\eea
The solutions of these conditions are either that $\lambda_1=\lambda_2=\lambda_3=0$ or $\lambda^2_1=\lambda^2_2=\lambda^2_3=1$ and $\lambda_1=\lambda_3 \lambda_2$.
In the former case the superalgebra is  $\mathfrak{osp}(6\vert 2)$ and in the latter case  a real form of $\mathfrak{sl}(3\vert 2)$ with even part $\mathfrak{sp}(2)\oplus \mathfrak{u}(3)$.
The choices of signs of $\lambda$'s denote the different embeddings of $\mathfrak{u}(3)$ in $\mathfrak{so}(6)$.  Again there is the possibility that the KSA is the non-simple
superalgebra  $\mathfrak{csl}(3\vert 2;7)$.  These results are in agreement with those in section 3.

\end{document}